\documentclass[12pt,amssymb]{article} 

\usepackage{mathrsfs}
\usepackage{amssymb}
\usepackage[dvips]{graphicx}

\newcommand{\newsection}[1]{
\addtocounter{section}{1} \setcounter{equation}{0}
\setcounter{subsection}{0} \addcontentsline{toc}{section}{\protect
\numberline{\arabic{section}}{{\rm #1}}} \vglue .6cm \pagebreak[3]
\noindent{ \bf  \thesection. #1}\nopagebreak[4]\par\vskip .3cm}
\newcommand{\newsubsection}[1]{
\addtocounter{subsection}{1}\setcounter{subsubsection}{0}
\addcontentsline{toc}{subsection}{\protect
\numberline{\arabic{section}.\arabic{subsection}}{#1}} \vglue .4cm
\pagebreak[3] \noindent{\it \thesubsection.
#1}\nopagebreak[4]\par\vskip .3cm}

\newcommand{\newsubsubsection}[1]{
\addtocounter{subsubsection}{1}
\addcontentsline{toc}{subsubsection}{\protect
\numberline{\arabic{section}.\arabic{subsection}.\arabic{subsubsection}}{
#1}} \vglue .4cm \pagebreak[3] \noindent{\it \thesubsubsection.
#1}\nopagebreak[4]\par\vskip .3cm}

\makeatletter
\newcommand{\seclabel}[1]{%
  \@bsphack
  \protected@write\@auxout{}%
     {\string\newlabel{#1}{{\thesection}{\thepage}}}
  \@esphack
  }
\newcommand{\subseclabel}[1]{%
  \@bsphack
  \protected@write\@auxout{}%
     {\string\newlabel{#1}{{\thesubsection}{\thepage}}}
  \@esphack
  }
\newcommand{\tablabel}[1]{%
  \@bsphack
  \protected@write\@auxout{}%
     {\string\newlabel{#1}{{\arabic{tabnum}}{\thepage}}}
  \@esphack
  }
\makeatother

\renewcommand{\theequation}{\thesection.\arabic{equation}}

\newlength{\extraspace}
\newlength{\extraspaces}
\newcounter{dummy}
\newcommand{\bc}{\begin{center}}
\newcommand{\ec}{\end{center}}
\newcommand{\be}{\begin{equation}
\addtolength{\abovedisplayskip}{\extraspaces}
\addtolength{\belowdisplayskip}{\extraspaces}
\addtolength{\abovedisplayshortskip}{\extraspace}
\addtolength{\belowdisplayshortskip}{\extraspace}}
\newcommand{\ee}{\end{equation}}

\newcommand{\ba}{\begin{eqnarray}
\addtolength{\abovedisplayskip}{\extraspaces}
\addtolength{\belowdisplayskip}{\extraspaces}
\addtolength{\abovedisplayshortskip}{\extraspace}
\addtolength{\belowdisplayshortskip}{\extraspace}}
\newcommand{\ea}{\end{eqnarray}}

\newcommand{\ban}{\begin{eqnarray*}
\addtolength{\abovedisplayskip}{\extraspaces}
\addtolength{\belowdisplayskip}{\extraspaces}
\addtolength{\abovedisplayshortskip}{\extraspace}
\addtolength{\belowdisplayshortskip}{\extraspace}}
\newcommand{\ean}{\end{eqnarray*}}
\newcommand{\baa}{
\addtocounter{equation}{1} \setcounter{dummy}{\value{equation}}
\setcounter{equation}{0}
\renewcommand{\theequation}{\thesection.\arabic{dummy}\alph{equation}}
\begin{eqnarray}
\addtolength{\abovedisplayskip}{\extraspaces}
\addtolength{\belowdisplayskip}{\extraspaces}
\addtolength{\abovedisplayshortskip}{\extraspace}
\addtolength{\belowdisplayshortskip}{\extraspace}}
\newcommand{\eaa}{
\end{eqnarray}
\setcounter{equation}{\value{dummy}}
\renewcommand{\theequation}{\thesection.\arabic{equation}}}

\input epsf

\newcounter{fignum}
\newcounter{tabel}

\newcounter{tabnum}
\newcommand{\vev}[1]{\left\langle #1\right\rangle}
\newcommand{\ket}[1]{\left| #1 \right\rangle}

\newcommand{\half}{\frac{1}{2}}
\newcommand{\del}{\partial}
\newcommand{\delb}{\bar{\del}}
\newcommand{\eol}{\nonumber \\}

\newcommand{\cO}{{\cal O}}
\newcommand{\Ext}{{\rm Ext}}

\newcommand{\bt}{{\bf 10}}
\newcommand{\bfv}{{\bf 5}}
\newcommand{\bfb}{{\overline{\bf 5 \!}\,}}
\newcommand{\btb}{{\overline{\bf 10 \!}\,}}

\begin{document}

%
%

\begin{flushright}
April 2009\\
AEI-2009-037
\end{flushright}
\vspace{2cm}

\thispagestyle{empty}

%
%

\begin{center}
{\Large\bf  Higgs Bundles and UV Completion in $F$-Theory
 \\[13mm] }

{\sc Ron Donagi}\\[2.5mm]
{\it Department of Mathematics, University of Pennsylvania \\
Philadelphia, PA 19104-6395, USA}\\[9mm]

{\sc Martijn Wijnholt}\\[2.5mm]
{\it Max Planck Institute (Albert Einstein Institute)\\
Am M\"uhlenberg 1 \\
D-14476 Potsdam-Golm, Germany }\\
[25mm]

 {\sc Abstract}

\end{center}

$F$-theory admits 7-branes with exceptional gauge symmetries, which
can be compactified to give phenomenological four-dimensional GUT
models. Here we study general supersymmetric compactifications of
eight-dimensional Yang-Mills theory. They are mathematically
described by meromorphic Higgs bundles, and therefore admit a
spectral cover description. This allows us to give a rigorous and
intrinsic construction of local models in $F$-theory. We use our
results to prove a no-go theorem showing that local $SU(5)$ models
with three generations do not exist for generic moduli. However we
show that three-generation models do exist on the Noether-Lefschetz
locus. We explain how $F$-theory models can be mapped to
non-perturbative orientifold models using a scaling limit proposed
by Sen. Further we address the construction of global models that do
not have heterotic duals. We show how one may obtain a contractible
worldvolume with a two-cycle not inherited from the bulk, a
necessary condition for implementing GUT breaking using fluxes. We
also show that the complex structure moduli in global models can be
arranged so that no dimension four or five proton decay can be
generated.

\newpage

\renewcommand{\Large}{\normalsize}

\tableofcontents

\newpage

\newsection{Introduction}

Recently \cite{Donagi:2008ca,Beasley:2008dc,Hayashi:2008ba}
initiated a systematic effort to study Kaluza-Klein GUT models in
$F$-theory. More precisely, we used an eight-dimensional gauge
theory with an exceptional gauge group, coupled to ten-dimensional
type IIb supergravity. The UV completion of this non-renormalizable
theory
is called $F$-theory \cite{Vafa:1996xn}. For practical purposes
however, very little is known about this non-perturbative
completion. We only know the low energy gauge theory and
supergravity Lagrangians, which are uniquely determined by the
symmetries. To get a reliable weakly coupled description in which
these Lagrangians can be trusted, the fields must be slowly varying.
Thus these models have a weakly coupled description in the large
volume limit, even though they are not in reach of perturbative
string theory. Recent work on $F$-theory models includes
\cite{Donagi:2008kj,Beasley:2008kw,Tatar:2008zj,Heckman:2008qa,
Heckman:2009bi,Hayashi:2009ge,Andreas:2009uf,
Blumenhagen:2008zz,Font:2008id,Blumenhagen:2008aw,Bourjaily:2009vf,Jiang:2008yf,
Collinucci:2008zs,Collinucci:2008pf,Heckman:2008es,Marsano:2008py}.

Despite the conceptual progress in
\cite{Donagi:2008ca,Beasley:2008dc,Hayashi:2008ba}, there were a
number of unanswered questions about the actual construction of
local models in $F$-theory. In particular, the strategy in
\cite{Donagi:2008ca} (and also \cite{Hayashi:2008ba}) relied on
taking a limiting form of models with a heterotic dual. This
approach yields manifestly consistent models, but it less than clear
if the most general local $F$-theory model is recovered this way.
The approach in \cite{Beasley:2008dc} is to postulate the matter
curves and the fluxes restricted to the matter curves. At first
sight this looks more flexible, but in this case it is less clear if
the data is mutually consistent. Given the uncertainties, it can be
hard to evaluate what $F$-theory does or does not predict.

The first purpose of this paper is to give a rigorous and intrinsic
construction of local $F$-theory models. The chain of logic is as
follows. As mentioned above, the basic idea is that we have to
construct compactifications of a supersymmetric eight-dimensional
gauge theory. Such compactifications are mathematically described by
meromorphic Higgs bundles. The main fact is that there is a natural
isomorphism between Higgs bundles and spectral covers in an
auxiliary non-compact Calabi-Yau geometry.\footnote{This observation
was made independently in \cite{Hayashi:2009ge}, which appeared
while this project was written up.} And the last set-up is the one
that allows us to make constructions, particularly of the fluxes.
Moreover these spectral covers are the same as one obtains from a
scaling limit of heterotic/$F$-theory duality. Thus, in a somewhat
roundabout way, our original strategy actually recovers all possible
local $F$-theory models. A completely parallel construction of local
$M$-theory models will appear in \cite{PW}.

Our spectral cover approach gives a precise description of the
configuration space of local $F$-theory models, which is important
for phenomenological applications. We will use this to classify the
possible matter curve configurations and prove a no-go theorem,
showing that the fluxes which were known to exist do not allow for a
local $SU(5)$ model with three generations. This is seen to imply
that in order to find realistic models, we have to solve a
Noether-Lefschetz problem, i.e. we have to tune the complex
structure moduli of a local model in order to find supersymmetric
solutions with three generations (which will then automatically have
stabilized some of the moduli). We then write down some new classes
of fluxes which are available on the Noether-Lefschetz locus, and
find the first examples of three-generation models. Such more
general fluxes are also available in heterotic models, where they
generally get mapped to {\it rigid} bundles. In fact we will point
out that heterotic constructions to date have been very special  and
essentially missed the landscape seen on the type II side. Along the
way we discuss several other interesting issues, such as orientifold
limits of $F$-theory models.

The second purpose of this paper is to begin the construction of
global UV completions which do not have a heterotic dual. {This
section was originally to appear as section 5 of
\cite{Donagi:2008kj}, but seemed to fit better with this paper.} We
will give some examples which should make the general strategy
clear. We do not find any meaningful constraints on extending
desired values of complex structure moduli from a local model to a
global model, thereby further validating the idea of studying local
models. In particular we find that it is possible to set the complex
structure moduli so that no dimension four or five proton decay can
be generated. But the understanding of global models is
unfortunately still rather incomplete. Our discussion focuses on
constructing compact models with desired 7-brane configurations, but
at present we do not have any good techniques for handling global
$G$-fluxes in general $F$-theory models. Constructing suitable
global fluxes is again an incarnation of a Noether-Lefschetz
problem, for which no really simple techniques seem to exist. One
possible approach using orientifold limits is briefly mentioned in
section \ref{IIblimits}.

\newpage

\newsection{Higgs bundles in $F$-theory}

In this section we will give a detailed description of local
$F$-theory models. Although much of this material is described
implicitly or explicitly in our previous papers, writing out the
chain of logic more carefully allows us to make sharper statements
about the configuration space of such models.

The reader should be aware that on occasion we use two different
definitions of the notion of a local model. The physical definition
is that of a model in which $M_{GUT}/M_{Pl}$ can be made
parametrically small. The other definition is that of a non-compact
$CY_4$ consisting of an $ALE$ fibration over a surface. Hopefully it
is clear from the context which notion we use.

\newsubsection{Local model from global model}

\subseclabel{LocalfromGlobal}

 Let us start with a global model,
which is defined as a compact elliptically fibered Calabi-Yau
complex four-fold with a section $\sigma(B_3)$ (often simply written
as $B_3$). The elliptic fibration can be described by a Weierstrass
model
\be\label{Weierstrass} y^2 = x^3 + fx + g \ee
where $f,g$ are sections of $K_{B_3}^{-4}, K_{B_3}^{-6}$
respectively. For the purpose of detecting singularities, it is more
useful to write the Weierstrass equation in generalized form as
\be\label{generalizedWeierstrass}
 y^2 + a_1 xy + a_3 y = x^3 + a_2 x^2 + a_4 x + a_6.
\ee
where the $a_i$ are sections of $K_{B_3}^{-i}$. By completing the
square and the cube, this may be written as (\ref{Weierstrass}), but
the generalized form is more convenient for prescribing singular
elliptic fibers along loci in $B_3$.

Suppose that we have a surface $S$ of singularities in $B_3$. This
will put certain restrictions on the sections $a_i$ above. Let us
take $z$ to be a coordinate on the normal bundle to $S$ in $B_3$, so
$S$ corresponds to $z=0$. We will often denote $c_1(NS) = -t$. Then
the order of vanishing of the $a_i$ may increase at $z=0$, so there
will be conditions of the form `$z$ divides $a_i$ at least $n_i$
times,' which are characteristic of the singularity type of the
elliptic fiber over $z=0$. These conditions have been worked out in
\cite{Tate,Bershadsky:1996nh} and are given in table \ref{Tate}
which was taken from \cite{Bershadsky:1996nh}. In retrospect, the
table is perhaps better understood in terms of Higgs bundles, which
we will discuss later. Now to get a local model from a global model,
we assign scaling dimensions to $(x,y,z)$ and drop the irrelevant
terms. Physically, this means we will be dropping certain higher
order terms in the $8d$ gauge theory.

\begin{figure}[p]
\thispagestyle{empty} \addtocounter{tabnum}{1} \tablabel{Tate}
\begin{center}
\renewcommand{\arraystretch}{1.35}
\begin{tabular}{|c|c|c|c|c|c|c|c|}
\hline

type  &  group  & \quad $ a_1$\quad  &
\quad $a_2$\quad  & \quad $a_3$ \quad & \quad $ a_4 $ \quad& \quad $ a_6$ \quad & $\Delta$ \\
\hline \hline $I_0 $  &  ---  & $ 0 $  & $ 0 $  & $ 0 $  & $ 0 $  &
$ 0$  & $0$ \\ $I_1 $  &  ---  & $0 $  & $ 0 $  & $ 1 $  & $ 1 $  &
$ 1 $  & $1$ \\ $I_2 $  & $SU(2)$  & $ 0 $  & $ 0 $  & $ 1 $  & $ 1
$  & $2$  & $
2 $ \\ $I_{3}^{ns} $  &  unconven.  & $0$  & $0$  & $2$  & $2$  & $3$  & $3$ \\
$I_{3}^{s}$  & unconven.  & $0$  & $1$  & $1$  & $2$  & $3$  & $3$ \\
$I_{2k}^{ns}$  & $ Sp(k)$  & $0$  & $0$  & $k$  & $k$  & $2k$  & $2k$ \\
$I_{2k}^{s}$  & $SU(2k)$  & $0$  & $1$  & $k$  & $k$  & $2k$  & $2k$ \\
$I_{2k+1}^{ns}$  & unconven.  &  $0$  & $0$  & $k+1$  & $k+1$  &
$2k+1$  & $2k+1$
\\ $I_{2k+1}^s$  & $SU(2k+1)$  & $0$  & $1$  & $k$  & $k+1$  & $2k+1$  & $2k+1$
\\ $II$  &  ---  & $1$  & $1$  & $1$  & $1$  & $1$  & $2$ \\ $III$  & $SU(2)$  & $1$
 & $1$  & $1$  & $1$  & $2$  & $3$ \\ $IV^{ns} $  & unconven.  & $1$  & $1$  & $1$
 & $2$  & $2$  & $4$ \\ $IV^{s}$  & $SU(3)$  & $1$  & $1$  & $1$  & $2$  & $3$  & $4$
\\ $I_0^{*\,ns} $  & $G_2$  & $1$  & $1$  & $2$  & $2$  & $3$  & $6$ \\
$I_0^{*\,ss}$  & $SO(7)$  & $1$  & $1$  & $2$  & $2$  & $4$  & $6$
\\ $I_0^{*\,s} $  & $SO(8)^*$  & $1$  & $1$  & $2$  & $2$  & $4$  &
$6$ \\ $I_{1}^{*\,ns}$
 & $SO(9)$  & $1$  & $1$  & $2$  & $3$  & $4$  & $7$ \\ $I_{1}^{*\,s}$  & $SO(10) $
 & $1$  & $1$  & $2$  & $3$  & $5$  & $7$ \\ $I_{2}^{*\,ns}$  & $SO(11)$  & $1$  & $1$
 & $3$  & $3$  & $5$  & $8$ \\ $I_{2}^{*\,s}$  & $SO(12)^*$  & $1$  & $1$  & $3$
 & $3$  & $5$ & $8$\\
$I_{2k-3}^{*\,ns}$  & $SO(4k+1)$  & $1$  & $1$  & $k$  & $k+1$
 & $2k$  & $2k+3$ \\ $I_{2k-3}^{*\,s}$  & $SO(4k+2)$  & $1$  & $1$  & $k$  & $k+1$
 & $2k+1$  & $2k+3$ \\ $I_{2k-2}^{*\,ns}$  & $SO(4k+3)$  & $1$  & $1$  & $k+1$
 & $k+1$  & $2k+1$  & $2k+4$ \\ $I_{2k-2}^{*\,s}$  & $SO(4k+4)^*$  & $1$  & $1$
 & $k+1$  & $k+1$  & $2k+1$
 & $2k+4$ \\ $IV^{*\,ns}$  & $F_4 $  & $1$  & $2$  & $2$  & $3$  & $4$
 & $8$\\ $IV^{*\,s} $  & $E_6$  & $1$  & $2$  & $2$  & $3$  & $5$  &  $8$\\
$III^{*} $  & $E_7$  & $1$  & $2$  & $3$  & $3$  & $5$  &  $9$\\
$II^{*} $
 & $E_8\,$  & $1$  & $2$  & $3$  & $4$  & $5$  &  $10$ \\
 non-min  &  ---  & $ 1$  & $2$  & $3$  & $4$  & $6$  & $12$ \\
\hline
\end{tabular}\\[5mm]
\parbox{14cm}
{ \bf Table \arabic{tabnum}: \it Results from Tate's algorithm
\cite{Tate,Bershadsky:1996nh}. The subscript s/ns stands for
split/non-split, meaning that there is/is not a monodromy action by
an outer automorphism on the vanishing cycles along the singular
locus. }
\renewcommand{\arraystretch}{1.0}
\end{center}
\end{figure}

For phenomenological purposes the case of most interest is a surface
$S$ of $I_5$ singular fibers. Then according to table \ref{Tate}, in
order to have an $SU(5)$ singularity along $z = 0$, we need the
leading terms near $z=0$ to be
\be a_1 = -b_5,\quad  a_2 = z b_4, \quad a_3 = -z^2 b_3, \quad a_4 =
z^3 b_2, \quad a_6 = z^5 b_0 \ee
where the $b_i$ are generically non-vanishing, and we may have
further subleading terms which vanish to higher order in $z$. The
$b_i$ are independent of $z$, so we may think of the $b_i$ as
sections of line bundles on the surface $S$. Now we assign scaling
dimensions $(1/3,1/2,1/5)$ to $(x,y,z)$ respectively. We throw out
the `irrelevant terms' whose scaling dimension is larger than one.
The resulting equation we get is
\be\label{localSU(5)model} y^2 = x^3 + b_0 z^5 + b_2 x z^3 + b_3 y
z^2 + b_4 x^2 z + b_5 xy \ee
which is exactly the equation of an $E_8$ singularity unfolded to an
$SU(5)$ singularity. The dimension one terms give the $E_8$
singularity and the terms with dimension smaller than one give a
relevant deformation of this singularity. Thus we may extract an ALE
fibration over $S$ from any global model by taking a suitable low
energy limit. Note that $c_1(B_3)|_S = c_1(S)-t$, and so the above
equation transforms as a section of $6c_1(S) - 6t$. Therefore the
Chern classes of the sections $b_i$ on $S$ are given by
\be b_i \sim (6-i) c_1(S) - t \ee

Note that we could have assigned different scaling dimensions to the
variables, which would result in dropping additional terms in
(\ref{localSU(5)model}). For instance if we assign degrees
$(1/3,1/2,2/9)$, then the $z^5$ term is also irrelevant and the
dimension one terms give the equation of an $E_7$ singularity.
However from the results of Tate's algorithm we see that it must
still be embedded in the $E_8$ singularity (\ref{localSU(5)model}),
so our choice will give the most general local model. The $E_8$
singularity is the maximal singularity that the elliptic fibration
allows without destroying the Calabi-Yau property.

Part of the attraction of local $F$-theory models is that almost all
of the observable sector is described by this one equation
(\ref{localSU(5)model}), plus a choice of $G$-fluxes. All the usual
complications of global models can be hidden in the subleading
corrections to this equation. This is equivalent to the statement
that the local geometry is completely described by the $8d$ gauge
theory. In the following we will analyze these local geometries in
more detail.

\newsubsection{Orientifold limits}

\subseclabel{IIblimits}

In this section, we analyze IIb limits of $F$-theory vacua. Such
limits are expected to be useful, since a number of issues
(particularly global issues) are currently much better understood in
the IIb theory than in $F$-theory. For instance we would like to use
this to analyse $G$-fluxes in global models. However as we will
discuss the regimes of validity are not overlapping and the IIb
models we get look very different from any previously considered IIb
GUT-like models. Thus there is still some work to be done to
understand the relation between the two pictures.

Consider again the Weierstrass equation
\be y^2 = x^3 + fx + g \ee
and its generalized form
\be y^2 + a_1 xy + a_3 y = x^3 + a_2 x^2 + a_4 x + a_6. \ee
As in \cite{Tate}, we define the following quantities:
\be
\begin{array}{rclrcl}
{\sf b}_2 &=& a_1^2 + 4 a_2 \qquad \qquad & {\sf b}_8 &=& 
{1\over 4} ({\sf b}_2 {\sf b}_6 - {\sf b}_4^2)\eol {\sf b}_4 &=& a_1
a_3 + 2 a_4 & \Delta &=& -{\sf b}_2^2 {\sf b}_8 -8{\sf b}_4^3 -27
{\sf b}_6^2 + 9 {\sf b}_2 {\sf b}_4 {\sf b}_6 \eol {\sf b}_6 &=&
a_3^2 + 4 a_6 & & & \eol \end{array} \ee
Then $f$ and $g$ may be recovered as
\ba\label{fginb} f &=& -{1\over 48}( {\sf b}_2^2 - 24 {\sf b}_4) \eol
g &=& -{1\over 864} (-{\sf b}_2^3 + 36 {\sf b}_2 {\sf b}_4 -216 {\sf b}_6)
\ea
Now supposed that we want to take a limit in the complex structure moduli space so that
the axio-dilaton becomes constant almost everywhere in the IIb space-time. Since
\be
j(\tau) = 4 {(24 f)^3\over 4 f^3 + 27 g^2 }
\ee
this will happen when
\be
{f^3 \over g^2 } \to {\rm constant}
\ee
Inspecting (\ref{fginb}), we see that the most evident way to
achieve this is by scaling up ${\sf b}_2$, or alternatively by
scaling down ${\sf b}_4$ and ${\sf b}_6$. Therefore let us consider
the following scaling limit:
\be
a_3 \to \epsilon \, a_3, \qquad a_4 \to \epsilon\,  a_4, \qquad a_6 \to \epsilon^2\,  a_6
\ee
Note that for our GUT models (\ref{localSU(5)model}), in this limit
$b_i/b_0$ scales like $1/\epsilon$ or $1/\epsilon^2$. Since $b_i/b_0
\sim {\rm Tr}(\Phi^i)$ are identified with Casimirs of the
eight-dimensional Higgs field, this means that the VEV of the Higgs
field is becoming large and we can no longer trust the $8d$ gauge
theory/$F$-theory description. One may still hope to get a different
weakly coupled description in terms of perturbative IIb string
theory. As we will discuss, this is possible, but we have to push
the model through a configuration with singularities that are
neither well-described by $F$-theory nor by perturbative type IIb.

Continuing, one finds
\ba f &=& -{1\over 48}( {\sf b}_2^2 - 24 \epsilon\,{\sf b}_4) \eol
g &=& -{1\over 864} (-{\sf b}_2^3 + 36 \epsilon\,{\sf b}_2 {\sf b}_4 -216\epsilon^2\, {\sf b}_6)
\ea
The discriminant is given by
\ba
\Delta &=& \epsilon^2(-{\sf b}_2^2 {\sf b}_8 -8\epsilon\,{\sf b}_4^3 -27\epsilon^2 {\sf b}_6^2
+ 9 \epsilon\,{\sf b}_2 {\sf b}_4 {\sf b}_6) \eol [1mm]
 &\sim &   -{1\over 4} \epsilon^2 \, {\sf b}_2^2({\sf b}_2 {\sf b}_6 -{\sf b}_4^2) + \cO(\epsilon^3) \ea
Therefore in the $\epsilon \to 0$ limit, all the roots are located
at ${\sf b}_2=0$ and ${\sf b}_2 {\sf b}_6 -{\sf b}_4^2=0$. The
monodromies around these roots were analyzed in
\cite{Sen:1997gv,Sen:1996vd}, with the result that
\be
 O7:{\sf b}_2=0, \qquad D7: {\sf b}_2 {\sf b}_6 -{\sf b}_4^2=0 \ee
Moreover, the $j$-function behaves as
\be j(\tau) \sim  {{\sf b}_2^4 \over \epsilon^2 \,({\sf b}_2 {\sf
b}_6 -{\sf b}_4^2)} \ee
which means that the string coupling goes to zero almost everywhere.
Therefore we get the following picture \cite{Sen:1997gv}: in the
limit of complex structure moduli space that we discussed above, the
Calabi-Yau four-fold becomes a constant elliptic fibration over a
Calabi-Yau three-fold given by
\be
\xi^2 = {\sf b}_2
\ee
where ${\sf b}_2 \sim K_{B_3}^{-2}, \xi \sim K_{B_3}^{-1}$. That is,
the emerging $CY_3$ is simply the double cover over $B_3$ with
branch locus given by ${\sf b}_2=0$. The orientifolding acts as
\be \xi \to -\xi, \quad y \to -y \ee
and the positions of the branes on this three-fold are given as
above. There are two copies of the $D7$ locus ${\sf b}_2 {\sf b}_6
-{\sf b}_4^2=0$ related by $\xi \to -\xi$.

Now let's apply this to our local models. The Calabi-Yau three-fold
will be given by a double cover of the total space of the normal bundle
$N_S \to S$, with branch locus given by ${\sf b}_2=0$. For $SU(5)$
models we get
\ba {\sf b}_2 &=& b_5^2 + 4 z b_4  \eol
{\sf b}_4 &=& z^2 b_3 b_5 + 2 z^3 b_2  \eol
{\sf b}_6 &=& z^4 b_3^2 + 4 z^5 b_0 \eol {\sf b}_2
{\sf b}_6 -{\sf b}_4^2 &=& z^5(4 b_3^2 b_4 -4 b_2 b_3 b_5 + 4 b_0
b_5^2 + z(16 b_0 b_4-4 b_2^2)) \ea
where $z$ is a local coordinate on the normal bundle $N_S$. Hence we
find a non-compact $O7$-plane along the branch locus ${\sf b}_2 =
0$, five gauge $D7$-branes wrapped on $S$, as well as a non-compact
flavour $D7$-brane. The $O7$-plane intersects the gauge 7-branes
along the matter curve
\be
\Sigma_\bt = \{ b_5 = 0 \}
\ee
which as expected carries an enhanced $SO(10)$ singularity. The
flavour $D7$-brane intersects the gauge $D7$-brane along
\be
\Sigma_\bfv = \{ R = b_3^2 b_4 - b_2 b_3 b_5 +  b_0 b_5^2 = 0 \}
\ee
which carries an enhanced $SU(6)$ singularity. Finally the Yukawa couplings
are localized at
\be
\lambda_{\rm top} \sim \{ b_5 = b_4 = 0 \},
\qquad \lambda_{\rm bottom} \sim \{ b_5 = b_3 = 0 \},
\ee
which carry enhanced  $E_6$ and $SO(12)$ singularities,
respectively.

Let us now look in more detail at the points of $E_6$ enhancement.
The equation of the Calabi-Yau can be written as
\be\label{upconifold} \xi^2 = u^2 + z w \ee
where $u = b_5$ and $w = 4 b_4$. Thus the $E_6$ points are conifold
singularities of the Calabi-Yau three-fold. We expect that the
limiting model has zero $B_{NS}$-field through the vanishing $S^2$,
so that it corresponds to a non-perturbative singularity of type
IIb.

Perturbative string theory breaks down at such conifold
singularities, and there are extra massless states. This should be a
chiral field corresponding to the zero modes of $B_2,C_2$ on the
`resolved' picture, or to a $D3$ wrapped on the vanishing $S^3$ in
the deformed picture.

In order to get a perturbative picture, we can try to resolve or
deform the conifold singularity. Let us first discuss the
resolutions. The two small ${\bf P}^1$'s are exchanged under the
discrete symmetry $\sigma: \xi \to -\xi$, and thus the small
resolution is projected out by the orientifold. The full orientifold
action is given by $\Omega (-1)^{F_L}\sigma$ where $\Omega$ is
worldsheet parity and $(-1)^{F_L}$ maps the RR fields to minus
themselves. The $NS$ $B$-field is odd under $\Omega (-1)^{F_L}$, so
it is consistent to have a non-zero value of $B$ through the
vanishing ${\bf P}^1$. So one can `resolve' the singularity by
turning on the $B$-field. ($C_2$ may also be non-zero; it is paired
with $B_2$ under SUSY). Thus there will be a description of the
up-type Yukawa coupling using $D1$-instantons. However there is no
smooth geometric picture, and $\alpha'$ corrections would be
important. The $B$-field may be tuned to the value $1/2$ which
corresponds to the quiver locus. These models are very different
from the IIb $SU(5)$ models that have been considered in the
literature (see eg. \cite{Blumenhagen:2008zz} for a recent
discussion and constructions), and more work needs to be done to
connect the two pictures.

We may also ask what happens with the flux that is responsible for
chiral matter in the scaling limit.
Likely this yields $U(1)$-flux for the overall $U(1)\subset U(5)$ in
the IIb model. In $F$-theory, this $U(1)$ becomes part of the larger
$E_8$ gauge symmetry, and is Higgsed by the adjoint field of the
$8d$ gauge theory.

Instead of trying to resolve the conifold points, one can also give
the $S^3$ a finite size by deforming the branch locus to a generic
section of $K_{B_3}^{-2}$. This is also compatible with the
orientifold action and removes the conifold points. (Three-form
fluxes through this $S^3$ are not compatible with the orientifold
action and can not be turned on). However this corresponds to
breaking the $SU(5)$ GUT group by giving an expectation value to a
field in the ${\bf 10}$. So although one could get a smooth
geometric background this way, it comes at the cost of breaking the
GUT group.

It is amusing to ask what happens for local $SO(10)$ models when we
take this limit. This corresponds to setting $b_5 \to 0$ identically
in the above equations. Then the $O7$-plane is reducible and
consists of a component wrapping $S$ and a component wrapped on the
curve $b_4 = 0$ in $S$ and stretching in the normal direction. The
spinors in the ${\bf 16}$ live on the intersection of the
non-compact orientifold plane with $S$ and are partially made of
non-perturbative $(p,q)$ strings.  The local equation of the
Calabi-Yau three-fold at these intersections is
\be \xi^2 = zw \ee
which means that they correspond to a curve of $A_1$ ALE
singularities. Presumably again $B_{NS}$ is zero here and they
correspond to non-perturbative singularities of type IIb; indeed
otherwise we would not expect massless modes of $(p,q)$ strings
here. Still this seems to be a very simple local model for producing
spinor representations in the IIb language.  The non-compact $D7$
brane intersects $S$ along two curves, one of which is the curve
above where the ${\bf 16}$ lives, and the other is $b_3=0$ which is
where the ${\bf 10}$ of $SO(10)$ lives.

Finally we can ask what happens for $E_6$ models. This corresponds
to setting both $b_5 \to 0$ and $b_4 \to 0$ identically in the above
equations. Then ${\sf b}_2$ vanishes identically so the limit we are
trying to take does not correspond to a IIb limit (except for very
special fibrations \cite{Dasgupta:1996ij}).

\newsubsection{Constraints from tadpole cancellation}

\subseclabel{Tadpoles}

From the local form of the singularity obtained above through the
results of Tate's algorithm, we may immediately deduce the homology
classes of the matter curves. Computing the discriminant of
(\ref{generalizedWeierstrass}), one finds
\be\label{localdiscriminant} \Delta = z^5 b_5^4(-{b_0} {b_5}^2
+{b_2} {b_3} {b_5}-{b_4}
   {b_3}^2) + \cO(z^6)\ee
Thus the matter curves are given by
\be \Sigma_\bt = \{ b_5 = 0\}, \qquad \Sigma_\bfv = \{ R = 0 \} \ee
which yields the following homology classes:
\be [\Sigma_\bt] = c_1-t, \qquad [\Sigma_\bfv] = 8c_1-3t \ee
In particular it follows that
\be\label{SU5tadpole} [\Sigma_\bfv] -3[\Sigma_\bt] - 5c_1 = 0 \ee
Of course we also know the precise equation of the matter curves,
but even these topological constraints are already quite
restrictive. Mathematically, these are necessary conditions for the
local geometry to be an elliptically fibered Calabi-Yau with
section.

Although it is clear from our construction that these constraints
have to be satisfied, it would be more satisfactory to give them a
physical interpretation. In six dimensional compactifications of
$F$-theory such constraints can be understood more physically as a
consequence of anomaly cancellation \cite{Sadov:1996zm}. For
instance the relation (\ref{SU5tadpole}) is then equivalent to
cancellation of the $tr_f(F^4)$ anomaly. One expects such relations
to hold also in more general $F$-theory settings
\cite{Tatar:2006dc}. We largely follow
\cite{Sadov:1996zm,Tatar:2006dc} in the remainder of this
subsection.

Consider the worldvolume of a 7-brane $S$, intersecting another
7-brane $S_a$ over a curve $\Sigma_a$. Under a gauge/Lorentz
transformation, in the presence of $(p,q)$ 7-branes we get an
additional contribution to the variation of the action given by
\be \delta_{\Lambda,\Theta} S \sim \int
I_{adj,6}^{1}(\Lambda,\Theta) \wedge \delta^2(S)\wedge \delta^2(S) -
\sum_{R_a} \int I^1_{R_a,6}(\Lambda,\Theta) \wedge
\delta^4(\Sigma_a) \ee
where $\Lambda$ is a local gauge transformation and $\Theta$ is a
local Lorentz transformation. Here $I^1_{R,6}$ is given through the
descent procedure as
\be dI^1 = \delta I^0, \quad dI^0 = I_{R,8} = \left[ {\bf
ch}_R(F)\wedge \hat{\bf A}(R)\right]_8 \ee
or more explicitly
\be \hat I_{R,8} = {1\over 24} {\rm Tr}_R (F^4) -{1\over 96} {\rm
Tr}_R(F^2) {\rm Tr}(R^2)+ {rk \over 128}\left({1\over 45} {\rm
Tr}(R^4)+{1\over 36} {\rm Tr} (R^2)^2 \right) \ee
where $F$ is understood to be the gauge field on the gauge 7-brane
wrapped on $S$, and $\hat I = (i(2\pi)^{d/2}) I$. Further we have
$\delta^2(S) \wedge \delta^2(S_a) = m_a\delta^4(\Sigma_a)$ and
$\delta^2(S) \wedge \delta^2(S) = -c_1(S) \wedge \delta^2(S)$. Note
that intersections are frequently not transverse in $F$-theory and
$m_a \not = 1$. This expression is the most straightforward
generalization of the usual expression for $D$-branes
\cite{Cheung:1997az,Minasian:1997mm}. The hypermultiplet spinors are
ordinary $6d$ spinors which do not carry $R$-charges, so the
expression for their anomaly is the usual one. The 8d gauginos also
carry $R$-charges which gives an extra contribution proportional to
$c_1(K_S)$. There could be further contributions to $\delta S$ in
compact models, but here we will concentrate on the pieces that are
associated to the gauge theory and have to be cancelled even in a
local model.

In order to check anomaly cancellation we convert all the gauge
traces to traces in the fundamental representation:
\be {\rm Tr}_R( F^4) = x_R\,{\rm Tr}_f (F^4) + y_R\, {\rm
Tr}_f(F^2)^2 , \qquad {\rm Tr}_R (F^2) = n_R\,{\rm  Tr}_f( F^2) \ee
In $F$-theory, the only massless tensor field available for the
Green-Schwarz mechanism is the RR field $C_4$.  Thus one would
expect that the anomaly can be cancelled by mediation of $C_4$ if
and only if the anomaly polynomial is factorizable, i.e. the matter
representations occurring are such that
\be \hat I_{12} =  \left[ \sum_{0,a} n_a \delta^2(S_a)\wedge(2{\rm
Tr}_f(F^2)-\half  {\rm Tr}(R^2) )\right]^2 , \ee
The corresponding tadpole cancellation condition is the well-known
constraint:
\be N_{D3} = {\chi(Y_4)\over 24} - {1\over 8\pi^2} \int_{Y_4} {\sf
G} \wedge {\sf G} \ee
Since all three terms receive unknown contributions from infinity,
we do not have to worry about this condition in a local model.

However this leaves a puzzle. The ${\rm Tr}_f(F^4)$ anomalies are
non-zero and localized at different places in the internal space. So
how do these pieces get cancelled exactly? There must be something
mediating them. In perturbative type IIb, the ${\rm Tr}_f( F^4)$ and
${\rm Tr}(R^4)$ anomalies on branes are cancelled by mediation of
the RR fields $C_0/C_8$. However in $F$-theory these fields are
massive and do not appear as propagating fields in the effective
action. Nevertheless it seems clear what must happen: in general
$F$-theory compactifications integrating out the massive modes of
the RR fields $C_0$ and $C_8$ leaves an effective interaction whose
variation cancels the ${\rm Tr}_f(F^4)$ anomalies.

A similar issue in fact also arises in $M$-theory on $G_2$ manifolds
and has been analyzed there \cite{Witten:2001uq} (see also \cite{PW}
for a discussion). In the $M$-theory setting, chiral fermions are
localized at points on the worldvolume of the gauge brane. In type
IIa the corresponding anomalies would be mediated by the RR gauge
field, but in $M$-theory this field is massive. Nevertheless there
is a residual interaction $\int K \wedge \omega^{(5)}$ which
transforms under gauge transformations, and the Gauss law for $K\sim
dA^{(1)}_{RR}$ is satisfied precisely when the ${\rm Tr}_f(F^3)$
anomalies are cancelled.

We have not precisely worked out the analogous statements in
$F$-theory. The problem is that if we apply the analogous trick,
rewriting $\int C_0 \wedge F^4 \sim - \int dC_0 \wedge \omega_7$, it
does not yield an interaction that is invariant under $Sl(2,Z)$
transformations, so it is incomplete. However for our purposes we
don't really need to work this out in detail, because we can use the
IIb orientifold limit identified in section \ref{IIblimits} to show
that the expected constraints have to be satisfied. In the IIb limit
the anomaly is cancelled by $C_0/C_8$ exchange as usual, and we get
the following modified Bianchi identity:
\be\label{F1tadpole} dF_1/2\pi = \sum_{D7} n_a \delta^2(S_a) - 8
\sum_{O7} \delta^2(O7) \ee
Here we use the `upstairs' picture, that is we write the relation on
the covering space before taking the orientifold quotient.
($F$-theory corresponds more naturally to the `downstairs' picture).

Now the integral of $dF_1$ over any closed two-cycle is zero. Let us
integrate over any curve $\Sigma_b$ in $S$, and let us write
(\ref{F1tadpole}) more suggestively as
\be dF_1/2\pi = 5 \delta^2(S) + \delta^2(S_a) + 5 \delta^2(S') - 8
\delta^2(O7) + {\rm other}\ee
where $S'$ is the mirror of $S$ under the orientifold action, the
$O7$-plane is the one intersecting $S$ over $\Sigma_\bt$ (where it
also intersects $S'$), and $S_a$ is the part of the $I_1$ locus
intersecting $S$ over $\Sigma_\bfv$. Then we find
\be 0 = -5 c_1(S) \cdot \Sigma_b + \Sigma_\bfv \cdot \Sigma_b +
(5-8)\Sigma_\bt \cdot \Sigma_b \ee
or equivalently
\be\label{SU(5)tadpole2}
 [\Sigma_\bfv] -3[\Sigma_\bt] - 5c_1 = 0  \ee
in $H_2(S, {\bf Z})$, which is what we wanted to show. More
generally we expect the relation
\be \sum_{R_a} x_{R_a} [\Sigma_a] -  \half x_{\rm adj}\,  c_1(S) = 0
\ee
to be equivalent to cancelling the ${\rm Tr}_f(F^4)$ anomalies, but
we have not been able to show this in full generality. As a special
case, in six-dimensional compactifications of $F$-theory the above
homology classes are all proportional to the class of a point, and
this relation was verified in \cite{Sadov:1996zm}.

Following \cite{Tatar:2006dc}, we may get a second constraint by
using a further relation in $F$-theory models:
\be \Delta = -12 K_{B_3} \ee
This is also a kind of 7-brane tadpole cancellation (eg. on $K3$ it
restricts the total number of 7-branes to be 24), but it
differs from (\ref{F1tadpole}). Since we have an
$SU(5)$ singularity along $S$, we may write
\be \Delta = 5[S] + \Delta' \ee
If we assume there are only matter curves for hypermultiplets in the
$\bfv$ or $\bt$, as is generically the case, then by intersecting
with $S$ we obtain
\be\label{7tadpole} -5 t + 4 \Sigma_\bt + \Sigma_\bfv = -12
K_{B_3}|_S \ee
Here we used $[S]\cdot [S] = c_1(NS)|_S = -t$.  The intersection
multiplicities can be read from the explicit form of the
discriminant (\ref{localdiscriminant}) (the coefficient of
$[\Sigma_\bt ]$ can presumably be understood from the fact that the
charge of an orientifold plane is $-4$  in the `downstairs'
picture). Further applying the adjunction formula $K_{B_3}|_S = K_S
+t$, we find that
\be 7\, t + 4 \Sigma_\bt + \Sigma_\bfv = -12 K_S  \ee
Together with the earlier constraint (\ref{SU(5)tadpole2}), it then
follows that the homology classes of the matter curves are given by
\be [\Sigma_\bt] = c_1(S) - t, \qquad [\Sigma_\bfv] = 8c_1-3t \ee
exactly as promised.

\newsubsection{Higgs bundles, spectral covers and ALE-fibrations}

\subseclabel{SpectralCover}

There are several equivalent descriptions of the supersymmetric
configurations of an $8d$ gauge theory. We may describe such a
configuration as an ALE fibration, which is how it arises in
$F$-theory in `closed string' variables. However we may also think
of it more intrinsically in terms of field configurations of the
adjoint scalars and gauge field. This gives us the Higgs bundle
picture. Finally we may replace the Higgs and gauge fields by their
eigenvalues. This gives us the spectral cover picture, or a fibered
weight diagram. The latter yields conventional $B$-branes in an
auxiliary non-compact Calabi-Yau three-fold $X$. The description of
$B$-branes in a Calabi-Yau is already a well-developed subject and
so this picture is the most convenient for doing actual
constructions and calculations. In this section, we spell out the
spectral cover description and its relation to the other pictures in
a bit more detail.

Much of the structure discussed here has been discussed in the
heterotic setting, but the main point is that it is in fact {\it
intrinsic} to the the $8d$ supersymmetric Yang-Mills theory and
therefore applies to an arbitrary local $F$-theory geometry, or any
other UV completion of $8d$ Yang-Mills theory. Moreover the spectral
cover description allows us to tie up some technical loose ends from
our previous papers. A completely analogous construction can be made
in $7d$ supersymmetric Yang-Mills theory \cite{PW} and leads to the
construction of local models in $M$-theory, in the large volume
limit where the Yang-Mills theory gives an accurate description. One
can also apply the dictionary for ALE fibrations over a Riemann
surface. This is essentially classic geometric engineering.

\newsubsubsection{The dictionary}

Given an ALE fiber over a point $p\in S$, we may choose a basis
$\alpha_i$ of $H_2(ALE_p, {\bf Z})$ corresponding to the fundamental
roots of the corresponding ADE Lie algebra (obviously this depends
on a choice of Weyl chamber). We may choose a dual basis $\omega^j$
of $H^2(ALE_p, {\bf Z})$ satisfying
\be \int_{\alpha_i} \omega^j = \delta^{ij} \ee
The Cartan generators for the adjoint fields arise
from deformations of the complex structure
\be \delta \Omega^{4,0} = \Phi^{2,0}_j \wedge \omega^j \ee
and the gauge fields arise from deformations of the three-form field
\be \delta C_3 = A_j \wedge \omega^j \ee
Further, the non-abelian generators arise from membranes wrapped on
the vanishing cycles of the ALE. Thus in $F$-theory, an ALE
fibration fibered over a surface $S$ yields precisely the data of a
supersymmetric $8d$ gauge theory compactified on $S$: a gauge field
$A$ on a bundle $E$ on $S$, and a `Higgs field' $\Phi$ which is a
section of
\be K_S \otimes Ad(G) \ee
where $G$ is the structure group of the bundle $E$.

The conditions for supersymmetry in the $8d$ gauge theory are
obtained by dimensional reduction. Namely we start with the
Hermitian-Yang-Mills equations in $10d$, and assume fields are
invariant under translation along a complex line. Then we can write
the gauge field as
\be {\sf A}^{0,1} = A_{\bar 1}(z^1,z^2) d\bar z^1 + A_{\bar
2}(z^1,z^2) d\bar z^2 + \Phi_{\bar 3}(z^1,z^2) d\bar z^3 \ee
The $F$-terms are
\be {\sf F}^{0,2}=0 \qquad\Rightarrow\qquad F^{0,2}=0, \qquad \bar
D_A \Phi = 0
 \ee
and the $D$-terms are
\be g^{i\bar j} {\sf F}_{i\bar j}= 0 \qquad \Rightarrow \qquad
g^{i\bar j} F_{i\bar j} + g^{i_1 \bar j_1}g^{i_2\bar
j_2}[\Phi^\dagger_{ \bar j_1 \bar j_2},\Phi_{ i_1 i_2} ] = 0 \ee
where $\Phi_{i_1 i_2}= \Phi_{\bar 3}\Omega_{ i_1  i_2  3}g^{3 \bar
3}$ is a $(2,0)$ form. The $D$-term is the moment map for gauge
transformations acting on the pair $(A,\Phi)$ with respect to the
K\"ahler form associated to the metric
\be g({\sf A},{\sf A})= \int |{\sf A}^{0,1}|^2 =  \int |A^{0,1}|^2 +
|\Phi^{2,0}|^2 \ee
The $F$- and $D$-term equations are called Hitchin's equations or
the Yang-Mills-Higgs equations. 
They are the critical points of two functionals, the holomorphic
Chern-Simons functional and the $D$-term potential:
\ba W &=& {1\over 4\pi} \int_S {\rm Tr} \,(A + \Phi) \delb (A+ \Phi)
+ {2\over 3}(A+\Phi)^3 \eol V_D &\sim& \half \int | J \wedge F +
[\Phi,\Phi^\dagger] |^2 \ea
In ALE fibrations, only the Cartan generators of $\Phi$ have a
non-vanishing VEV, and so we have $[\Phi,\Phi^\dagger]=0$. Such
solutions are said to be regular. Field configurations with
$[\Phi,\Phi^\dagger]\not =0$ should have an alternative description
as space-filling 9-branes satisfying the Hermitian Yang-Mills
equations. In the following we will assume that $[\Phi,\Phi^\dagger]
=0$. Then this data defines a Higgs bundle
\cite{HitchinHiggs,Simpson}.

Since $[\Phi,\Phi^\dagger] =0$, the real and imaginary parts of
$\Phi$ can be simultaneously diagonalized and we may try to replace
$\Phi$ by its spectral data, i.e. its eigenvalues and eigenvectors.
For convenience we temporarily focus on $SU(n)$ gauge groups, though
analogous constructions exist for any gauge group. We let $s$ denote
a coordinate on the canonical bundle $K_S$. The Hitchin map is the
map that sends the Higgs field $\Phi$ to its Casimirs. In the
$SU(n)$ case, the Casimirs are the coefficients of the polynomial
\be\label{Hitchinmap} \det(sI - \Phi) = 0 \ee
This polynomial equation makes sense globally on a non-compact CY
three-fold $X$, consisting of the total space of the canonical
bundle $K_S \to S$. For a generic point on $S$ the roots $\lambda_i$
of this polynomial give us $n$ points on the fiber of $K_S$. Thus
the $n$ roots trace out a complex surface $C$ which covers the zero
section $n$ times. This is the spectral cover for the fundamental
representation of $SU(n)$. Since we will be interested in
non-compact covers, we should allow simple poles for the Higgs
fields. We can get rid of the poles in (\ref{Hitchinmap}) by
multiplying with a suitable section. Thus instead of
(\ref{Hitchinmap}) we will write the degree $n$ equation
\be 0 = b_0 s^n+ b_1 s^{n-1} + b_2 s^{n-2} + \ldots + b_n   \ee
Since $s=0$ is marked, the only coordinate transformations allowed
are rescaling. For $SU(n)$ gauge groups we further want to impose
that all the roots add up to zero. Since we have
\be \lambda_1 + \ldots + \lambda_n = b_{1},\ee
therefore we set $b_{1} = 0$. The surface $C$ is non-compact. Along
the locus $b_0=0$, two of the roots go off to infinity. Let us
denote the divisor $b_0=0$ on $S$ by $\eta$. Since $s$ is a
coordinate on $K_S$, the $b_i$ are then seen to be sections of
\be b_i \sim \eta  - i\, c_1(S) \ee

We further have to describe the gauge field $A$ in this picture, or
equivalently the bundle $E$. To do this, it is useful to think of
$\Phi$ as a map
\be \Phi: E \to E \otimes K_S \ee
Then under the action of $\Phi$, each fiber of $E$ can be decomposed
into its eigenspaces $\oplus_i {\bf C}\ket{i}$. Let us denote
coordinates on the total space $K_S \to S$ by pairs $(p,s)$ where
$p\in S$ and $s$ is the coordinate on the fiber. The assignment
\be (p,\lambda_i) \to {\bf C}\ket{i} \ee
yields a line bundle $L$ on $C$ called the spectral line
bundle.%
\footnote{More precisely, let us denote $R = K_{C/S}$ the
ramification divisor, and $s \in H^0(C,p_C^*K_S)$ the tautological
eigenvalue section, whose value at a point $(p,s)$ is given by $s$.
Then $L\otimes \cO(-R)$ is the kernel of $p_C^* \Phi - s I:p_C^*E
\to
p_C^*E \otimes p_C^*K_S$.} %
Furthermore since $\bar D_A\Phi = 0$, $\bar D_A$ commutes with the
action of $\Phi$ on $E$, and we can we can simultaneously
diagonalize $\bar D_A$. Thus we get a holomorphic connection on $L$,
and we can pick a section $\ket{i} \in {\bf C}\ket{i}$ by parallel
transport. Note that the spectral cover and line bundle in $K_S$
satisfy the usual requirements of a $B$-brane in the large volume
limit: a holomorphic cycle with a holomorphic bundle on it.

Conversely, given a spectral cover and a spectral line bundle, we
may recover the Higgs field $\Phi$ and the bundle $E$. We may
represent $\Phi$ as
\be \Phi = \sum_i \lambda_i \Pi_i \ee
where $\Pi_i$ is the projection on ${\bf C}\ket{i}$. More formally
we can pull-back $\Phi$ to the total space of the canonical bundle.
Then we may write it as the canonical section
\be \pi^*\Phi(p,s) = sI \ee
where $I$ is the identity operator. Therefore given the spectral
data, we may recover the Higgs bundle as:
\be E = p_{C*}L, \qquad \Phi = p_{C*} s \ee
The gauge field $A$ is obtained as the push-forward of a connection
on $L$. Furthermore if we want an $SU(n)$ bundle rather than a
$U(n)$ bundle, then we also need to require
\be \det(p_{C*}L) = \cO \ee
where $\cO$ is the trivial line bundle. This gives a topological
constraint on the allowed spectral line bundles.

We may go back and forth between this description and the ALE
fibration. For $SU(n)$ gauge groups the $A_{n-1}$-ALE fibration is
defined by the following equation
\be\label{AnALE} y^2 = x^2 + b_0 s^n+  b_2 s^{n-2} + \ldots + b_n
\ee
As far as the variation of Hodge structure is concerned, the
quadratic terms $x^2$ and $y^2$ are irrelevant and may be dropped,
recovering our previous equation. This argument is well known from
Landau-Ginzburg models, where we `integrate out' the fields with a
quadratic potential.

Furthermore in terms of the ALE fibration $Y_4$, the spectral line
bundle is encoded as $G$-flux. Let us think of (\ref{AnALE}) as a
conic bundle fibered over the complex plane parametrized by $s$. We
have a map
\be p_R : R \to C \ee
where $R$ is obtained from $C$ by attaching a line (with equation
$y=x$) to each point in the fiber of the covering $C \to S$.
Furthermore we have a map
\be i: R \to Y_4 \ee
which embeds these lines in the ALE (\ref{AnALE}), each line sitting
at the corresponding point $s=\lambda_i$ in the $s$-plane. Let us
decompose the flux of the spectral line bundle as
\be c_1(L) = \half c_1(K_{C/S}) + \gamma \ee
where $K_{C/S} = K_C - p_C^* K_S$ is the ramification divisor, and
$p_{C*} \gamma = 0$. Then the spectral line bundle and the $G$-flux
are related by
\be G = i_*p_R^*\gamma - q\, P_{Y_4}(S) \ \in H^{2,2}(Y_4) \ee
Here $P_{Y_4}(S)$ is the Poincar\'e dual to the zero section $S$ in
$Y_4$ (which can be dually represented by an ALE fiber), and $q$ is
determined by requiring that $\int_S G = 0$, which gives $q = \gamma
\cdot_C \Sigma_E$. Given this explicit expression it is not too hard
to check that such fluxes are always primitive, i.e. satisfy $J
\wedge G = 0$ on $Y_4$, if $p_{C*}\gamma = 0$. For $U(n)$ bundles,
we need to make sure that if $J$ contains a piece $\pi^*J_{S}$
pulled-back from $S$, then $p_{C*}\gamma\cdot J_S = 0$.

We may state this more naturally as follows. We have a single charge
lattice $\Lambda=\oplus_i {\bf Z}\ket{i}$ varying over $S$, which
can be given two equivalent interpretations. In the ALE picture
$\Lambda$ is identified with $H_2(ALE,{\bf Z})$. In the spectral
cover picture it is identified with $\oplus_i {\bf Z} e_i$, where
$e_i$ are the nodes of the corresponding ADE Dynkin diagram.
Similarly the dual lattice $\Lambda^*$, which is actually isomorphic
to $\Lambda$ because ADE lattices are self-dual, is given by
$H^2(ALE,{\bf Z})$ in the ALE picture, or alternatively by $\oplus_i
{\bf Z} e_i^*$. These two local systems are naturally isomorphic.
The $b_i/b_0$'s correspond to the invariant polynomials of a
meromorphic section of $\Lambda^* \otimes K_S$. The flux of the
spectral line bundle or equivalently the $G$-flux corresponds to a
generator of
\be   H^2(S,\Lambda^*)\cap H^{1,1}(S, \Lambda^* \otimes {\bf C}) \ee

\newsubsubsection{Other associated spectral covers}

The spectral cover we have considered so far should really be called
$C_{E}$, to indicate that it corresponds to the fundamental
representation. We can also construct spectral covers for other
representations, which typically describe equivalent data. One
important cover that we will need is the spectral cover
$C_{\Lambda^2E}$ for the anti-symmetric representation of $SU(n)$.
This has $\half n (n-1)$ sheets. Each sheet intersects a fiber of
$K_S$ in the points
\be \Lambda^2E: \qquad \lambda_i + \lambda_j, \quad i<j \ee
where addition is defined in the obvious way in each fiber. In fact
it is not hard to write down an explicit equation using mathematica.
For the case $n=5$, the cover is defined by the degree 10 equation
\ba 0&=&s^{10} + 3 s^8 c_{2} - s^7 c_{3} + s^6 (3 c_{2}^2 - 3 c_{4})
+
 s^5 (-2 c_{2} c_{3} + 11 c_{5})+
  s^4 (c_{2}^3 - c_{3}^2 - 2 c_{2} c_{4}) \eol
  & &  +
 s^3 (-c_{2}^2 c_{3} + 4 c_{3} c_{4} + 4 c_{2} c_{5}) +
 s^2 (-c_{2} c_{3}^2 + c_{2}^2 c_{4} - 4 c_{4}^2 + 7 c_{3} c_{5})
 \eol
 & & +s (c_{3}^3 + c_{2}^2 c_{5}
 - 4 c_{4} c_{5})- c_{3}^2 c_{4}+ c_{2} c_{3} c_{5} - c_{5}^2
 \ea
where $c_i = b_i/b_0$ and the whole equation should be multiplied
with $b_0^3$ in order to remove the denominators. We denote the
intersection of $C_{\Lambda^2E}$ with the zero
section $s=0$ by $\Sigma_{\Lambda^2E}$.%
\footnote{Note that the subscript here indicates the representation
of the holonomy group, not the unbroken gauge group. In our
discussion later however we will instead use the subscript to denote
the representation under the GUT group, as in our previous papers.
Thus in our $SU(5)$ examples later we will have $\Sigma_{\Lambda^2E}
= \Sigma_\bfv$ and $\Sigma_{E} = \Sigma_\bt$.}
The surface $C_{\Lambda^2E}$ is singular when two of the eigenvalues
coincide, i.e. $\lambda_i+\lambda_j = \lambda_k + \lambda_l$ for
some $i,j,k,l$. This happens in codimension one, so the matter curve
$\Sigma_{\Lambda^2E}$ is also singular at isolated points. The
spectral line bundle on this cover is given fiberwise by
\be L_{\Lambda^2E}:\qquad (p,\lambda_i+\lambda_j) \to {\bf
C}\ket{i}\wedge \ket{j} \ee
It is not really a line bundle but a (torsion-free) sheaf, its rank
jumping up at the singular locus, and one has to desingularize in
order to define things unambiguously. Still this data is determined
uniquely by the spectral line bundle for the cover of the
fundamental representation, as follows.

In order to write an unambiguous formula it is more natural to think
about unembedded covers \cite{DSpectral}. We take pairs of points
$(q_1,q_2) \in C_E \times_S C_E$, and remove the diagonal where $q_1
= q_2$. Then we define the quotient\footnote{Strictly we have to
take the closure and then take the quotient. We oversimplified this
issue here and in the remainder in order to avoid too much
notation.}
\be \tilde C_{\Lambda^2E} = \{(q_1,q_2) \in C_{E}\times_S C_{E}\ |\
 \ q_1\not = q_2 \}/{\bf Z}_2\ee
where the ${\bf Z}_2$ action interchanges $(q_1,q_2) \to (q_2,q_1)$.
This cover is embedded in $X\times_S X/{\bf Z}_2$, but not in $X$,
and provides a resolution of $C_{\Lambda^2E}$. There is a natural
map
\be  C_{E} \times_S  C_{E} - {\rm diag}(C_E) \quad \to\quad \tilde
C_{\Lambda^2E}\quad \to\quad C_{\Lambda^2E} \ee
The last map is given fiberwise by sending $(\lambda_i, \lambda_j)
\to \lambda_i + \lambda_j$. The pairs $(\lambda_i, \lambda_j)$ and
$(\lambda_k, \lambda_l)$ are distinct in $\tilde C_{\Lambda^2E}$
even when  $\lambda_i+\lambda_j = \lambda_k + \lambda_l$ in
$C_{\Lambda^2E}$. The inverse image of $\Sigma_{\Lambda^2E}$ in
$\tilde C_{\Lambda^2E}$ is its normalization $\tilde
\Sigma_{\Lambda^2E}$. The spectral line bundle $L_E$ on $C_{E}$ gets
mapped to a smooth line bundle on $\tilde C_{\Lambda^2E}$:
\be L_{E} \times L_{E} \quad \to\quad \tilde L_{\Lambda^2E}\quad
\to\quad L_{\Lambda^2E}\ee
It only gets mapped to a sheaf $L_{\Lambda^2E}$ on $C_{\Lambda^2E}$
because the map $\tilde C_{\Lambda^2E} \to C_{\Lambda^2E}$ is
two-to-one at the singular locus, but this is irrelevant since we
should work with the non-singular surface $\tilde C_{\Lambda^2E}$.
This construction should be interpreted as follows. The spectral
line bundle on $C_{\Lambda^2E}$ is the set of eigenlines
$\ket{i}\wedge \ket{j}$ of $\Lambda^2E$ under the action of the
Higgs field. When $\lambda_i+\lambda_j = \lambda_k + \lambda_l$ the
cover $ C_{\Lambda^2 E}$ is singular, so there is an ambiguity in
assigning eigenlines of $\Lambda^2E$ to eigenvalues of $
\Phi_{\Lambda^2 E}$ in a neighbourhood of the singular locus. This
ambiguity is naturally resolved by recalling that the assignment of
eigenlines to eigenvalues was unambiguous for $E$ (assuming $C_E$ is
smooth), in other words it is naturally resolved by requiring that
$L_{\Lambda^2E}$ descends from a smooth line bundle on $\tilde
C_{\Lambda^2E}$. As emphasized in \cite{Hayashi:2008ba}, this means
that keeping track of the gauge indices implies that the
hypermultiplet at the intersection really couples to $\tilde
L_{\Lambda^2E}$. Thus the hypermultiplet propagates on the
normalized matter curve $\tilde \Sigma_{\Lambda^2E}$ rather than on
$\Sigma_{\Lambda^2E}$ itself.

Similarly we may construct spectral covers for other
representations. For instance the spectral cover for the symmetric
representation $C_{S^2E}$ is given fiberwise by
\be S^2E: \qquad \lambda_i + \lambda_j, \quad i\leq j \ee
We will not have any need for these other coverings in this paper.

\newsubsubsection{Fermion zero modes}

Now that we have a description of configurations in the $8d$ gauge
theory in terms of holomorphic cycles and bundles on them, we would
like to describe the zero modes of the Dirac operator. In
holomorphic geometry the Dirac operator splits into a Dolbeault
operator
\be \bar{D} = \delb + A^{0,1} + \Phi^{2,0}_\Omega \ee
and its adjoint $\bar{D}^\dagger$. Here $\Phi^{2,0}_\Omega$ is our
$\Phi^{2,0}$ Higgs field contracted with the anti-holomorphic
$(0,3)$ form on the non-compact Calabi-Yau $X$, yielding a $(0,1)$
form whose index lies in the normal direction to $S$ in $X$. Since
supersymmetry is preserved $\Omega$ is covariantly constant and
hence the normal bundle is identified with the canonical bundle. The
spinor configuration space together with the $\bar{D}$ operator
yield a complex whose cohomology is computed by $\Ext$-groups. On
the other hand, the zero modes of $\not \! D = \bar{D} +
\bar{D}^\dagger$ are also in one-one correspondence with the
cohomology of $\bar{D}$. Let us denote by $i,j$ the embedding of
divisors into $X$, and assume that $R,R'$ are sheaves on these
divisors. Then the wave functions of fermion zero modes are in
one-one correspondence with generators of $\Ext$ groups:
\be \Ext^p_X(i_*R, j_*R') \ee
As usual, the index $p$ correlates with the $4d$ chirality as
$(-1)^p$. For $p=1,2$ the $4d$ part of the wave function belongs to
a chiral (anti-chiral) superfield, and for $p=0,3$ we get
four-dimensional gauginos (or possibly ghosts if suitable stability
conditions are not satisfied). These cohomology groups are of course
localized on the intersection of the supports.

The $\Ext$-groups naturally give a unified description of all the
possibilities. Let us assume that the full spectral cover splits up
into a multiple of the zero section (the `gauge brane') and some
additional non-compact pieces (the `flavour branes'). If we assume
that $i$ embeds the zero section in $X$ and $j$ embeds the remainder
of the spectral cover in $X$, then this reduces to
\be \Ext^p_X(i_*R, j_*R') \sim H^{p-1}(\Sigma, R^\dagger \otimes R'
\otimes K_S|_\Sigma) \ee
for the case of intersecting branes, and
\be \Ext^p_X(i_*R, i_*R') \sim H^{p-1}(S,R^\dagger \otimes R'\otimes
K_S) \ \oplus\  H^p(S,R^\dagger \otimes R') \ee
for the case of coincident branes, as was deduced in
\cite{Donagi:2008ca,Beasley:2008dc}. To be more precise, there is a
`spectral sequence' which starts with the right-hand-side and may
possibly lift some of the zero modes to arrive at the
left-hand-side, as was actually noted in \cite{Donagi:2008ca}. In
typical examples this lifting does not happen, and so we can take
this relation to be an equality. The number of moduli of the
configuration is given by
\be N_{mod} = \Ext^1_X(j_*L_E,j_*L_E) \sim h^{2,0}(C) \oplus
h^{0,1}(C). \ee
Here we assumed that the spectral cover is smooth. Then the number
of moduli is independent of the spectral line bundle on $C$, which
just cancels in the formula. Similarly the number of adjoints is
given by
\be N_{adj} = \Ext^1_X(i_*\cO_S,i_*\cO_S) \sim h^{2,0}(S) \oplus
h^{0,1}(S) \ee
Further, the unambiguous formula for the amount of chiral matter on
$\tilde \Sigma_{\Lambda^2E}$ is given by
\be\label{Lambda2matter} \Ext^1(i_*\cO_S, j_*\nu_*\tilde
L_{\Lambda^2E})\ \sim \ H^0(\tilde \Sigma_{\Lambda^2E},\tilde
L_{\Lambda^2E}\otimes \nu^*K_S|_{\tilde \Sigma_{\Lambda^2E}}) \ee
where $\nu$ is the normalization $\tilde \Sigma_{\Lambda^2E} \to
\Sigma_{\Lambda^2E}$. This recovers the answer found in
\cite{Hayashi:2008ba}.

The spectral cover description also allows us to give a precise
mathematical definition of the classical Yukawa couplings (and
higher dimension couplings as well), at least up to field
redefinitions. It is simply given by the Yoneda pairing:
\be \Ext^{p_1}(i_{1*}R_1,i_{2*}R_2) \times
\Ext^{p_2}(i_{2*}R_2,i_{3*}R_3) \times
\Ext^{3-p_1-p_2}(i_{3*}R_3,i_{1*}R_1) \to {\bf C} \ee
Again this expression summarizes all the possibilities, with wave
functions either localized in the bulk or on 7-brane intersections.
One should be careful about drawing conclusions from such
computations however. The usual warnings about the relation with the
physical Yukawa couplings (which depend on the K\"ahler potential
and may receive loop corrections) apply.

\newsubsubsection{$E_8$ Higgs bundles}

Now we return to the case of primary interest. In local $F$-theory
models we are dealing with fibrations by $E_8$ ALE spaces, or
equivalently with $E_8$ Higgs bundles. The relevant spectral cover
is the one for the adjoint representation, which we will simply call
`the' spectral cover. The adjoint representation is $248$
dimensional, of which eight are Cartan generators. Thus the full
spectral cover will have $248$ sheets. In order to break to an
$SU(5)$ GUT group, we turn on an $Sl(5,{\bf C})$ Higgs bundle. The
adjoint representation of $E_8$ decomposes as
\be {\bf 248} = ({\bf 24},{\bf 1}) + ({\bf 1},{\bf 24}) + (\bfv,\bt)
+ (\bfb,\btb) + (\bt,\bfb) + (\btb,\bfv) \ee
Thus the $E_8$ spectral cover breaks up into several pieces, which
can be labelled by representations of the holonomy group of the
Higgs bundle. Clearly the relevant spectral covers are those for the
fundamental representation and for the anti-symmetric representation
of $SU(5)$.

 \begin{figure}[t]
\begin{center}
            \scalebox{.45}{
               \includegraphics[width=\textwidth]{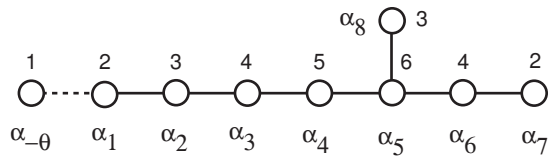}
               }
\end{center}
\vspace{-.5cm} \caption{ \it The extended $E_8$ Dynkin diagram and
Dynkin indices.}\label{E8Dynkin}
\end{figure} 

Referring back to the general form of  a local $SU(5)$ model as
derived using Tate's algorithm:
\be\label{localSU(5)again} y^2 = x^3 + b_0 z^5 + b_2 x z^3 + b_3 y
z^2 + b_4 x^2 z + b_5 xy \ee
The parameters can be identified with the following Casimirs of a
meromorphic $Sl(5,{\bf C})$ Higgs bundle:
\be C_i(\Phi) \sim {\rm Tr}(\Phi^i) \sim b_i/b_0 \ee
To see this, the singularity (\ref{localSU(5)again}) is generically
of type $A_4$, but by sequentially tuning the $b_i$ to zero we get
successively $SO(10),E_6,E_7$ and an $E_8$ singularity. Since the
holonomy group of the Higgs bundle is the commutant of the gauge
group in $E_8$, then the parameters must correspond to the indicated
Casimirs. (A more precise way to see this \cite{Donagi:2008ca} is by
using the $F$-theory/heterotic duality map). We see that there
exists a canonical map between the parameters in the ALE fibration,
and an $SU(5)$ spectral cover in $K_S \to S$ defined by
\be b_0 s^5 + b_2 s^2 + \ldots + b_5  = 0 \ee
Note that $\eta$ is related to our earlier $t$ by $\eta = 6c_1 - t$.
The five roots $\{\lambda_1, \ldots, \lambda_5\}$ of this polynomial
determine the sizes of all the cycles of the $E_8$ ALE space. Recall
that
\be
\Phi\ket{i} = \lambda_i \ket{i}
\ee
As discussed in section 5.1 of \cite{Donagi:2008kj}, we can take the
five roots of the polynomial to correspond to the periods of the
following cycles (up to Weyl permutations)
\be
\begin{array}{rclrcl}
 \ket{1} &=& \alpha_4  & \ket{4} &=& \alpha_1 +\alpha_2 + \alpha_3 + \alpha_4  \eol
\ket{2} &=& \alpha_3 +\alpha_4 & \ket{5}
&=& \alpha_{-\theta} + \alpha_1 + \alpha_2 + \alpha_3 +
\alpha_4 \eol
\ket{3} &=& \alpha_2 + \alpha_3 + \alpha_4 \qquad  & & &
\end{array}
 \ee
The sizes of the cycles $\{\alpha_5, \ldots ,\alpha_8\}$ are taken to be zero,
generating an $SU(5)$ GUT group, and all other cycles are obtained as linear combinations.
The matter curve $\Sigma_\bt$
corresponds to $\lambda_i=0$ for some $i$, the matter curve
$\Sigma_\bfv$ corresponds to $\lambda_i + \lambda_j=0$ for some
$i,j$, etc. The top Yukawa is localized at $\lambda_i=\lambda_j=
\lambda_i+\lambda_j=0$, the bottom at
$\lambda_i+\lambda_j=\lambda_k+\lambda_l =
\epsilon_{ijklm}\lambda_m=0$, and the $\bfv\cdot\bfb\cdot{\bf 1}$ at
$\lambda_i+\lambda_j=\lambda_j+\lambda_k=\lambda_i-\lambda_k=0$.

\newsubsection{Construction of fluxes}
\subseclabel{InheritedFluxes}

Let us briefly recap what we saw above. Local models in $F$-theory
correspond to ALE fibrations over a surface $S$ with $G$-flux.
Physically we expect that this data, the ALE fibration and the
$G$-flux, can be described as configurations in an $8d$
supersymmetric Yang-Mills theory compactified on $S$, i.e. a Higgs
bundle. This is indeed the case, and moreover this data is also
equivalent to a covering $C_E$ of the zero section in an auxiliary
non-compact Calabi-Yau three-fold $X = (K_S \to S)$, together with a
holomorphic line bundle. In a IIb-like language, we can call this
covering a non-compact flavour brane, whose intersection with the
gauge brane (which is wrapped on the zero section of $X$) yields the
matter curve $\Sigma_\bt$. The group theory of $E_8$ implies that
there is a second flavour-brane $C_{\Lambda^2E}$, completely
determined by the first covering, whose intersection with the gauge
brane yields the matter curve $\Sigma_\bfv$.

Constructing the branes, or equivalently the ALE fibration, is easy:
we only need to specify the $b_i$, which are sections of line
bundles on $S$ with Chern classes given by $\eta - i c_1$. In order
to get chiral matter however we must actually turn on a flux on
$C_E$, which will determine a unique flux on $C_{\Lambda^2E}$ by
group theory. In this subsection we discuss the issue of
constructing such fluxes.

In order to facilitate the analysis we will compactify the local
Calabi-Yau to
\be \bar X ={\bf P}(\cO \oplus K_S) \ee
$\bar X$ is certainly not a Calabi-Yau; the Calabi-Yau metric
diverges at infinity.  We denote by $\cO(1)$ the line bundle on
$\bar X$ which restricts to the eponymous line bundle on each ${\bf
P}^1$-fiber. We may choose homogeneous coordinates $(u_1,u_2)$ on
the ${\bf P}^1$-bundle which are sections of $\cO(1)$ and $\cO(1)
\otimes K_S$ respectively. The coordinate $s$ used previously is
identified with $u_2/u_1$.

The spectral cover $C\subset X$ is compactified to a compact surface
$\bar{C}\subset \bar X$ by adding a divisor $\eta_\infty$ at
infinity. The equation
\be b_0s^5  + b_2 s^3 + b_3 s^2 + b_4 s  + b_5 = 0 \ee
has a double zero at $u_1 = 0$. Therefore $\eta_\infty$ covers
$\eta$ exactly once and is isomorphic to it. We denote the
cohomology class dual to the zero section (the Poincar\'e dual of
$S$ in $\bar X$) by $s_0$ and the class of the section at infinity
by $s_\infty=c_1(\cO(1))$. Then we have $s_\infty = s_0 + c_1(TS)$
and $0 = s_0 \cdot s_\infty = s_0\cdot (s_0 + c_1(TS))$.

We would like to lift line bundles on $C$ to line bundles on the
compact surface $\bar{C}$ which are easier to study. If the genus of
$\eta_\infty$ is non-zero, then there may be line bundles on $C$
that cannot be lifted to $\bar C$. However any algebraic line bundle
on $C$ can be lifted. To see this, any algebraic line bundle on $C$
is of the form $\cO(D)_C$ for some divisor $D$. Let $\bar D$ be the
closure of $D$ in $\bar C$. Then $\cO(\bar D)_{\bar C}$ gives a lift
of $\cO(D)_C$ as desired. Moreover global $G$-fluxes are algebraic
and yield an algebraic class in the local model. Therefore from here
on we may restrict our attention to extendable line bundles.

Now consider a spectral line bundle $L$ on $\bar C$. The
corresponding Higgs bundle is given by $E =p_{C*}L$, $\Phi = p_{C*}
s$. We have
\be c_1(p_{C*}L) = p_{C*}c_1(L) -\half p_{C*}r  \ee
where $r$ is the ramification divisor, $r = -c_1(\bar C) +
p_C^*c_1(S)$. Explicitly we find that
\be r = 
(n-2)s_0 + p_C^*(\eta -c_1(S)) \ee
If $c_1(p_{C*}L)$ is not zero, then we have a $Gl(n,{\bf C})$ Higgs
bundle rather than an $Sl(n,{\bf C})$ Higgs bundle. For
phenomenological applications we want the latter, so we need to
impose the restriction $c_1(p_{C*}L)=0$ on the allowed spectral line
bundles. Then it is convenient to decompose
\be c_1(L) = \half r + \lambda \gamma \ee
where $\lambda$ is a parameter. The condition $c_1(p_{C*}L)=0$ is
then equivalent to $p_{C*}\gamma = 0$. The class $r/2$ is generally
not integer quantized. Since $c_1(L)$ must be integer quantized,
$\gamma$ must compensate and can generally not be an integer class
either, but it will always be a rational linear combination of
integer classes.

So our task is to find integer classes $\gamma$ with $p_{C*}\gamma =
0$ in $H^2(\bar C,{\bf Z})$. To get a supersymmetric configuration,
$\gamma$ must further be of Hodge type $(1,1)$. As we will now
argue, for generic complex structure moduli there exists only one
such class (up to multiplication by an integer), and we can write it
down explicitly.

As for line bundles on $\bar{C}$, we can use the Lefschetz-Noether
theorem. $\bar{C}$ is the zero locus of a section of $\cO(n) \otimes
L_{\eta-nc_1}$, which is usually an ample line bundle since $\cO(n)$
is ample and $L_{\eta-nc_1}$ is effective and non-zero. Therefore
there is an injective map $i^*:H^{1,1}(\bar{X}) \to
H^{1,1}(\bar{C})$. As a result, $H^{1,1}(\bar{C})$ splits into two
pieces, the classes inherited from $\bar{X}$ and the primitive
classes. The Noether-Lefschetz theorem says that when $\bar{C}$ is
ample, then for `generic' complex structure moduli there are no
primitive classes in $H^{1,1}(\bar{C})$.

Let us write down the inherited class explicitly. The cohomology
group $H^2(\bar X, {\bf Z})$ is spanned by $s_0$ and $\pi^*H^2(S,
{\bf Z})$, i.e. the pull-back of classes on $S$ to $\bar X$.
Therefore the inherited classes in $H^{1,1}(\bar C)$ are spanned by
the class of the matter curve $\Sigma_\bt$, as well as any class on
$S$ pulled back to $\bar{C}$. (In particular, $c_1(\bar C)$ is in
this span, by the adjunction formula, and so is $\eta_\infty$). Our
class $\gamma$ will be a linear combination of those, but it also
needs to satisfy $p_{C*}\gamma =0$, which is clearly not satisfied
by any class pulled back from $S$. Thus we can single out the class
$[\Sigma_E]$ and subtract the `trace', i.e. we single out the
following unique linear combination:
\be \gamma_u = n[\Sigma_E]- p_C^*p_{C*}[\Sigma_E] = n[\Sigma_E]-
p_C^*(\eta - n c_1) \ee
We used the subscript $u$ on $\gamma_u$ to indicate that this class
is universal, i.e. it always exist in a local model. In the last
equality we just the fact that $p_{C*}[\Sigma_E]$ is just the class
$[\Sigma_E]$ sitting inside $H^2(S)$, and since it is given by
${b_n}=0$ it follows that it can also be written as $\eta - n c_1$.

Let us define a line bundle using this class. Its first Chern class
will be given by
\be c_1(L) = \half r + \lambda \gamma_u \ee
with $\gamma_u$ as defined above and $\lambda$ a parameter. From our
explicit expressions for $r$ and $\gamma_u$, we see that $c_1(L)$ is
an integer class when $\lambda$ is an integer and $n$ is even, or
when $\lambda=\half+$ integer and $n$ is odd. For this corresponding
choice of spectral line bundle, we can deduce the net amount of
chiral matter. It is given by
\be N_{\rm chiral} =  -\chi(i_*\cO_S, j_*L) = +\chi(L\otimes
K_S|_{\Sigma_\bt}) = \lambda \int_{\Sigma} \gamma_u  \ee
where in the last equality we used the Riemann-Roch formula and the
fact that $( r/2+c_1(K_S))|_\Sigma = -c_1(\Sigma)/2$.
We have
\be \Sigma \cdot_{\bar C} \Sigma = S_0 \cdot_{\bar X} S_0
\cdot_{\bar X} \bar C = -c_1(S_0) \cdot_{S_0}  \Sigma \ee
Further we have $\Sigma \cdot_{\bar C} p^* \alpha = \alpha
\cdot_{S_0} \Sigma$ for any $\alpha \in H^2(S,{\bf Z})$. Applying
this with $\alpha = \eta - n c_1$, we see that
\be \gamma_u \cdot_{\bar C} \Sigma = -\eta \cdot_{S_0} \Sigma \ee
Therefore we find
\be  N_{\rm chiral} = \lambda \int_{\Sigma} \gamma_u  =  -\lambda
\eta (\eta - nc_1) \ee
This is of course the same formula as encountered in spectral cover
constructions in the heterotic string \cite{Curio:1998vu}.

Therefore the only fluxes available for general complex structure
moduli will give the conventional chirality formula known from the
heterotic string. We do not see more general options in the local
$F$-theory set-up. We will call such fluxes {\it inherited} or {\it
universal}. However, there do exist more general fluxes, both in the
heterotic setting and in the $F$-theory setting. The point is that
general fluxes are not supersymmetric for generic Higgs bundle
moduli, and thus are not among the fluxes that we found above. For
special values of the moduli (which is called the Noether-Lefschetz
locus) there are additional supersymmetric fluxes available, and
turning on such fluxes would therefore automatically stabilize some
of the moduli. We will call such fluxes primitive or {\it
non-inherited}. Generic fluxes are non-inherited. They exist for
both $F$-theory spectral covers and heterotic spectral covers, where
they give rise to rigid bundles on the $CY_3$ after Fourier-Mukai
transform. But they are harder to write down and have not really
been analyzed in either context.

\newsubsection{Further constraints}

\subseclabel{NoGo}

In the previous sections we encountered a number of constraints that
must be satisfied for consistency of the local model. Now we would
like to consider imposing a few further constraints, that are not
needed for consistency but are likely needed to get a realistic and
calculable four-dimensional model. We will concentrate on $SU(5)$
models, so there is a matter curve
\be [\Sigma_\bt] = c_1-t \ee
which must be effective and non-zero.

The K\"ahler class $J$ is a generator of $H^2(B_3, {\bf R})$ that
must be positive on all the effective cycles of the geometry. Modulo
these positivity constraints, there is an independent scale in the
geometry for every generator of $H^2(B_3, {\bf R})$. In order to get
a model that is calculable and predictive, we need some small
parameters that we can expand in. The main requirement that we want
to make is that $M_{GUT}/M_{Pl}$ is unbounded from below, where
$M_{GUT} \sim V_S^{1/4}$ and $M_{Pl} \sim V_{B_3}^{1/6}$. Now it is
a priori possible that in a given model we can take $V_{B_3} \to
\infty$ while keeping $V_S$ finite, therefore decoupling the GUT and
Planck scales, but we cannot take $V_S\to 0$ while keeping $V_{B_3}$
finite. However this would depend on the geometry of $B_3$, and
moreover would normally leave additional scales in the model with
physics that cannot be decoupled from the visible sector. Needless
to say that would not be beneficial for the predictiveness of the
model.\footnote{Perhaps an exception would be if $B_3$ is fibered
over $S$, but this can be excluded by condition (3) below.} To get a
predictive model in which the visible sector is largely independent
of the rest of $B_3$, we will require that one can take $V_S\to 0$
while keeping $V_{B_3}$ (or any other cycles not inside $S$) finite.
Moreover this yields a local constraint that can be checked without
knowing the compactification manifold $B_3$.

There are two ways in which we could take $V_{S} \to 0$ while
keeping other cycles fixed. The first is that we could require $S$
to contract to a point. This will turn out to be a very strong
condition which will essentially single out a unique model. We could
also require $S$ to contract to a curve of singularities. This is a
less stringent condition, but together with some other physical
constraints will still rule out a good deal of models.

Thus our first assumption is as follows:
\begin{enumerate}
\item{\it Contractibility.} $S$ can be contracted to a point.
By Grauert's criterion \cite{Grauert}, this means that the class $t$
must be ample, in particular $t\cdot C > 0 $ for any curve $C$ in
$S$.\footnote{Even though we have seen that for many purposes $S$
can be regarded as living inside the total space of the canonical
bundle, this criterion has nothing to do
with contractibility in the auxiliary local Calabi-Yau. %
We will see this more explicitly in the global examples later.}

\end{enumerate}
We can draw some immediate conclusions from this assumption. Since
$c_1 -t$ is effective and non-zero, and $t$ is ample, $c_1$ must be
effective and non-zero. Therefore $K_S^{-n}$ cannot have sections
for any positive $n$ and the Kodaira dimension is $-\infty$. From
the classification of surfaces, we then know that $S$ is related to
${\bf P}^2$ or a ruled surface (i.e. a ${\bf P}^1$-fibration over a
Riemann surface of genus $g$) by a sequence of blow-ups and
blow-downs.

The ruled surfaces have $h^{1,0}(S) = g$, which would lead to
massless adjoint fields in the effective four-dimensional theory if
$g>0$. This looks phenomenologically undesirable, so we will exclude
this possibility with our second assumption:
\begin{enumerate}
\item[2.]{\it No adjoint scalars.} The Hodge numbers of $S$ must satisfy
$h^{0,1}(S) =0$ and $h^{2,0}(S)=0$.
\end{enumerate}
Then $S$ is either ${\bf P}^2$ or can be obtained by a sequence of
blow-ups from a Hirzebruch surface ${\bf F}_r$. Note this includes
all the del Pezzo surfaces. The divisors on ${\bf F}_r$ are
generated by $b,f$ and $E_i$, with the intersection numbers
\be b\cdot b = -r, \qquad b \cdot f = 1, \qquad f\cdot f = 0, \qquad
b\cdot E_i = f \cdot E_i = 0, \qquad E_i \cdot E_j = -\delta_{ij}
\ee
By exchanging $b$ and $b+f$, we may take $r\geq 0$. Further we have
\be c_1({\bf F}_r) = 2b + (r+2) f -\sum_{i=1}^k E_i \ee
Similarly we may write
\be t = n_b b + n_f f - \sum_{i=1}^k n_i E_i \ee
Let us first assume we are on ${\bf F}_r$, with no blow-ups. From
ampleness of $t$ we get $n_b>0, -n_b r + n_f > 0$. Since $c_1-t$ is
effective and non-zero, we also get $n_b \leq 2$ and $n_f\leq r+2$,
with strict inequality for $n_f$ if $n_b=2$ or vice versa. Then we
either have $n_b = 1$ and $r<n_f\leq r+2$, or we have $n_b = 2$,
$n_f = r+1$ and $r = 0$ or $1$.

We may add a further reasonable assumption which eliminates most of
these models. Currently, there is only one known mechanism for
breaking the GUT group while preserving the standard GUT relations
at leading order \cite{Donagi:2008kj,Beasley:2008kw}. This mechanism
requires a $-2$-class on $S$ (i.e. a class with $x\cdot x = -2$) in
order to avoid massless lepto-quarks. This class must further be
orthogonal to any classes that are inherited from $B_3$ in order to
avoid Higgsing hypercharge or loosing the standard $SU(5)$ relations
between the gauge couplings at leading order. Let us take this as
our third assumption:
\begin{enumerate}
  \item[3.]{{\it GUT breaking using fluxes.}} There must exist a $-2$-class
$x \in H^2(S, {\bf Z})$ which is orthogonal to any class inherited
from $H^2(B_3, {\bf Z})$. In particular, $x \cdot c_1 = x \cdot t =
0$

\end{enumerate}
Let us again consider the Hirzebruch surfaces. Then $h^2({\bf F}_r)
= 2$ so by condition (3) it follows that $t$ must be a rational
multiple $a c_1$ of $c_1$. Since $t \cdot f > 0$, the coefficient
$a$ must be positive, and since $c_1 - t$ must be effective, the
coefficient $a$ must be $\leq 1$. But this happens only for $r$ even
in which case $c_1$ is divisible by $2$, so this leaves
\be S={\bf F}_r \quad  {\rm with} \ r \ {\rm  even}, \qquad
\Sigma_\bt = \half c_1 \ee
But now by condition (1) we get $t\cdot b = -r+2>0$, so this leaves
only $S = {\bf F}_0$ and $t = \half c_1$. Note that ${\bf P}^2$ is
also ruled out by condition (3).

Now we consider the case of Hirzebruch surfaces with at least one
blow-up. Again we have the constraints above from $t\cdot b>0$,
$t\cdot f>0$ and $c_1-t$ effective. However we also get $t \cdot E_i
= n_i > 0$ and $t\cdot (f-E_i) = n_b - n_i>0$. Hence we must have
\be t = 2b + (r+1)f -\sum_{i=1}^k E_i \ee
From $t\cdot b =-r+1>0$ we find that $r=0$. Moreover, ${\bf F}_r$
with one blow-up is actually the same surface as ${\bf F}_1$ with
one blow-up, so $r=0$ is ruled out as well, and therefore all cases
with blow-ups are ruled out. So we conclude that assumptions (1)-(3)
leave a unique possibility for $S$ and $t$:
\be\label{goodmodel} (1) + (2)+(3) \ \Rightarrow \quad S = {\bf
F}_0, \qquad t = \half c_1, \qquad \Sigma_\bt = \half c_1 \ee
We will study this case in more detail later in the paper. In
particular we will show how to engineer three-generation models and
how to embed it in a global model.

It is evident by now that condition (1) in particular is quite
strong. In order to have $V_S \to 0$ while keeping other cycles
fixed, we can also replace assumption (1) by: 
\begin{enumerate}
  \item[1'.]{\it Contractibility.} $S$ can be contracted to a curve,
  i.e. $S$ admits a fibration $F \to S \to B$ where the fibers $F$
  can be contracted to a curve $B$ of singularities.

\end{enumerate}
In this case $t$ is not necessarily ample, but $t$ must be ample
when restricted to the components that are being contracted
\cite{Ancona,Peternell}. Therefore, $t\cdot C> 0$ when $C$ is the
general fiber $F$ or an irreducible component of the singular
fibers.

A priori the base $B$ of the fibration can have any genus $g$, in
which case we would have $h^{1,0}(S) = g$. Again by assumption (2)
the base $B$ is restricted to be ${\bf P}^1$. Likewise, the fibers
must be rational: the curve $c_1-t = \Sigma_{\bt}$ is effective and
the fiber $F$ moves, so there must be some copy of $F$ that is not
contained in $\Sigma_{\bt}$. Therefore we must have $(c_1-t)\cdot
F\geq 0$. Since $F$ gets contracted, we have $t\cdot F
> 0$, and it follows that $c_1 \cdot F > 0$. But by the adjunction formula we
have $c_1 \cdot F = 2 - 2g(F)$, hence $c_1 \cdot F = 2$ and $F$ is a
${\bf P}^1$ as promised.

So our $S$ is a ruled surface with rational base and fibers. From
the classification of surfaces, we know that $S$ is rational and can
be obtained by blowing up some points on a Hirzebruch surface ${\bf
F}_r$. The most general possibility is obtained by blowing up some
points on a conic bundle, which allows for the possibility of a
multiple fiber. Our argument below is actually even more
constraining when there are multiple fibers, so in what follows we
will focus on the case that none of the fibers is multiple.

If $S$ is a Hirzebruch surface then we can run the previous
argument. Using condition (3) it follows that $t=\half c_1$ and $r$
is even. Apart from these we must consider possible blow-ups of
${\bf F}_r$. Again we write
\be t = n_b b + n_f f - \sum_{i=1}^k n_i E_i \ee
Under assumption (1') we can no longer conclude that $t\cdot b$ must
be positive, but we still know that $t$ must be positive on $f, E_i$
and $f-E_i$. From $t \cdot f>0$, $t \cdot E_i>0$ and $t\cdot
(f-E_i)>0$ we get $n_b>0$, $n_i > 0$, and $n_b - n_i
> 0$. From $c_1-t$ effective and non-zero we get  $n_b \leq 2$
and $n_f \leq r+2$, with strict inequality for $n_f$ if $n_b=2$ and
$n_i = 1$. Therefore the only possibility is
\be S = B_k({\bf F}_r), \qquad t = 2b + n_f f - \sum_{i=1}^k E_i ,
\qquad \Sigma_\bt = (r+2-n_f) f \ee
with $n_f < r+2$. Here we used $B_k$ to denote blowing-up $k$ times.
These possibilities also satisfy condition (3), since there are
classes of the form $f-E_i-E_j$ and $E_i-E_j$ which are orthogonal
to $c_1$ and $t$. Moreover we can't do too many blow-ups. Recall
that the sections $b_i$ specifying an $SU(5)$ model live in $c_1 -
t, \ldots,6c_1 -t$, so these line bundles need to admit sections.

So we conclude that under assumptions (1'), (2) and (3), we get the
following possibilities for $S$ and $t$:
\be\label{Frmodels} S={\bf F}_r \quad  {\rm with} \ r \ {\rm  even},
\qquad t = \half c_1, \qquad \Sigma_\bt = \half c_1 \ee
or

\be\label{BkFr} S = B_k({\bf F}_r), \qquad t = 2b + n_f f -
\sum_{i=1}^k E_i , \qquad n_f < r+2.\ee
The remaining possibility $S = {\bf P}^2$ is still ruled out by
assumption (3).

 We may consider adding one final assumption. We will soon see
though that this assumption has an important loophole, so it will be
weakened significantly.
\begin{enumerate}
  \item[4.]{{\it Three generations.}} The net number of generations is given by
\be -\lambda (6c_1-t)\cdot_{dP} (c_1-t)  \ee
where $\lambda \in {\bf Z} + \half$.
As we argued, this is the only universal formula one can write down.
However this does not represent the most general configuration of
local $F$-theory models and will be revisited in section
\ref{NoninheritedFluxes}.

\end{enumerate}
Let's apply this to all the possibilities we found. For the
Hirzebruch surfaces with $t = \half c_1$ we find that the minimal
number of generations is eleven. For the blow-ups of Hirzebruch
surfaces with $t$ as in (\ref{BkFr}), we find that the minimal
number of generations is $5 \times (r+2-n_f)$, and in general it is
always divisible by $5$. We conclude it is not possible to make a
local three generation $SU(5)$ model under these assumptions.

If we drop condition ($1$) or ($1'$) it is not hard to find
three-generation models. For instance, the $dP_8$ example in
\cite{Donagi:2008ca} with $\eta = 6 c_1(S)$ is consistent and
satisfies assumptions (2) and (3), but it does not satisfy
assumption ($1$) or ($1'$) since it has $t=0$, and is therefore not
a truly local model.

It may be interesting to point out that the three generation
$SO(10)$ models in \cite{Donagi:2008ca} (which have $\eta = 4c_1 +
E$ where $E$ is any $-1$-curve, and $\Sigma_{\bf 16} = [E]$) do
satisfy the conditions (1),(2) and (3) for $dP_k$ with $2\leq k \leq
7$. However a fully satisfactory way of breaking the $SO(10)$ GUT
group in these models while preserving gauge coupling unification
has not yet been identified.

To summarize this subsection, under conditions (1) -- (3) we only
found one possibility for $S$ and $t$, listed in (\ref{goodmodel}).
Under assumptions (1') -- (3) we only found the possibilities listed
in (\ref{Frmodels}) and (\ref{BkFr}). Using the inherited fluxes
(assumption (4)), none of these models could account for three
generations. In the following subsections, we will examine some
possible loopholes in our assumptions.

\newsubsection{Another way to break the GUT group}
\subseclabel{NewGUTBreaking}

In \cite{Donagi:2008kj,Beasley:2008kw} the GUT group was broken to
the Standard Model gauge group by turning on an abelian flux. A
priori there exists a second possibility: one may also break the GUT
group to the Standard Model by turning on an abelian Higgs field.

To do this, we take a degree six spectral cover
\be b_0 s^6 + b_2 s^4 + \ldots + b_6 = 0 \ee
which generically breaks $E_8$ to $SU(3) \times SU(2)$. Note that
$b_1$ must vanish if the structure group is to be in $SU(6)$ rather
than $U(6)$. Now if the $b_i$ are such that this equation factorizes
\be (c_0 s + c_1)(d_0 s^5 + d_1 s^4 + \ldots +d_5) = 0 \ee
where $c_0 d_1 + c_1 d_0 = 0$, then the structure group of the Higgs
field commutes with $SU(3) \times SU(2) \times U(1)$.

The matter curves are easy to find. Consider first an irreducible
degree 6 spectral cover (this was worked out in
\cite{Hayashi:2008ba}). One uses the following decomposition of the
adjoint of $E_8$ under $SU(3) \times SU(2) \times SU(6)$:
\ba {\bf 248} &=& ({\bf 8,1,1}) + ({\bf 1,3,1}) + ({\bf 1,1,35})
\eol & & + ({\bf 3,2,6}) + ({\bf \bar 3,2,\bar 6}) +({\bf 3,1,15}) +
({\bf \bar 3 ,1,\bar 15}) + ({\bf 1,2,20}) . \ea
We have the following matter curves:
\ba ({\bf 3,2}) &\to& \{b_6=0\}  \eol ({\bf 3,1}) &\to & \{b_0 b_5^3
-b_2b_3b_5^2 + b_4 b_5 b_3^2-b_3^3 b_6 =0\}\eol ({\bf 1,2}) &\to &
\{ b_6(b_2^2 - 4 b_4 b_0) + b_0 b_5^2-b_2b_5b_3+b_4b_3^2 =0\} \ea
In the reducible case we simply substitute the $b_i$ for the
appropriate bilinears in $c_i$ and $d_j$.


In addition there can be hypercharged scalars. To see this, recall
that the moduli of the spectral cover are counted by
\be \Ext^1_X(i_*L,i_*L) \ee
When the spectral cover is reducible, this decomposes as
\be \sum_{m,n} \Ext^1_X(i_{m*}L_m,i_{n*}L_n) \ee
where $m,n$ run over the irreducible components. The off-diagonal
zero modes are clearly charged under the extra $U(1)$'s since their
VEVs smooth the spectral cover and break these $U(1)$'s. These modes
are localized at the intersection $c_0 s + c_1 =d_0 s^5 + d_1 s^4 +
\ldots +d_5$. In order to get a realistic model, such hypercharged
scalars must obtain a mass, i.e. we must obstruct the deformation of
two irreducible components of the cover into a single smooth piece.
This could be done by turning on spectral line bundles on the
irreducible components that do not arise as the limit of a line
bundle on the smooth deformation. Possibly some other mechanism like
an instanton effect can also be used.

There is unfortunately one issue with this scenario. The polynomial
$b_6$ is a section of $N_S$. Thus not only $c_1 -t$ should have a
section, but also $-t$. By our classification of the possible pairs
$(S,t)$, even with condition (3) dropped, such an $S$ can not be
contractible, which means it is not a true local model. Thus this
mechanism can only be used if we drop the requirement that
$M_{GUT}/M_{Pl}$ can be made parametrically small.

A second concern is that there is a potential $D$-term instability.
The $U(1)$ will be non-anomalous if we succeeded in lifting the
charged moduli mentioned above, but there may be a bare FI term.
This FI term is given by the moment map
\be \zeta \sim G \wedge J \sim  F \wedge J \ee
where $F$ is the flux of the spectral line bundle/sheaf on the
degenerate spectral cover. We've argued that spectral line bundles
on smooth spectral covers with $c_1(p_{C*}L)=0$ are primitive. This
will also hold for line bundles on degenerate spectral covers if
they are obtained by taking a limit of a line bundle on a smooth
spectral cover, because primitiveness is a closed condition. However
these are precisely not the line bundles we want, because they would
not lift the hypercharged scalars. Hence on degenerate spectral
covers one would need to look at specific models in more detail.

\newsubsection{Non-inherited fluxes}
\subseclabel{NoninheritedFluxes}

The results of section \ref{NoGo} clarify our options. Dropping the
first assumption is a priori possible and leads to consistent
models, but it would mean we can not make an expansion in
$M_{GUT}/M_{Pl}$ and therefore would greatly diminish the
predictiveness of $F$-theory GUTs. Dropping assumption (3) means
that we need an alternative mechanism to break the GUT group while
preserving the standard $SU(5)$ relations at leading order. We made
such an alternative proposal in section \ref{NewGUTBreaking} but it
did not seem to be compatible with contractibility of $S$. Therefore
we are led to drop assumption (4) and investigate the possibilities
of non-inherited $G$-fluxes.

The argument in section \ref{NoGo} that the inherited fluxes are not
sufficient by no means rules out three generation models. Rather it
means that we need to look at more general fluxes that are not
critical points of the holomorphic Chern-Simons superpotential for
generic moduli, and we have to do more work to show that there
exists a stable supersymmetric minimum. In mathematics circles this
would be called a Noether-Lefschetz problem. On the other hand, the
first three assumptions already ruled out all but a handful of
7-brane configurations, listed in (\ref{goodmodel}),
(\ref{Frmodels}) and (\ref{BkFr}) in section \ref{NoGo}. Thus in
contrast to eg. heterotic model building, we have a very restricted
set of possibilities to start with and we know all the continuous
parameters.

Let us first ignore the requirement of supersymmetry, and simply ask
if there are any fluxes available, not necessarily of type $(1,1)$,
which might give three generations. This will be the case if we can
show there exists a class $\gamma_2 \in H^2(\bar C, {\bf Z})$ which
is orthogonal to $p_C^*H^2(S,{\bf Z})$ and satisfies $\gamma_2 \cdot
\Sigma_\bt = 1$, since then we can add some integer multiple of it
to $\half r + \half\gamma_u$ and get any number of generations we
want. Equivalently, $\gamma_2$ must satisfy
\be \gamma_2 \cdot \gamma_u = \gamma_2 \cdot (5\Sigma_\bt -
p_C^*p_{C*} \Sigma_\bt) = 5 \ee
Now the lattice $H^2(\bar C, {\bf Z})$ modulo torsion is unimodular
by Poincar\'e duality. Thus if $\gamma_u$ is primitive in the sense
that it is not an integer multiple of a smaller integer class, then
there exists an $\alpha \in H^2(\bar C, {\bf Z})$ such that
\be \alpha \cdot \gamma_u = 1 \ee
Then defining $\gamma_2 = 5 \alpha - p_C^*p_{C*}\alpha$, we have
$\gamma_2 \cdot \gamma_u = 5$ and $p_{C*}\gamma_2 = 0$ as required.
Therefore we are guaranteed that the required fluxes exist if
$\gamma_u$ is primitive in the sense above. In a unimodular lattice,
a sufficient condition is that $\gamma_u \cdot \gamma_u$ is not
divisible by any square. It is easy to see that $\gamma_u \cdot
\gamma_u = 5 \gamma_u \cdot \Sigma_\bt$ which we have already
computed in section \ref{NoGo}. For $S={\bf P}^1 \times {\bf P}^1$
with $t = \half c_1$, it is equal to $5 \times 22$ which does not
have any squares. So we conclude from a purely topological argument
that it is possible to obtain three generations, although this
argument cannot establish that there is a supersymmetric minimum for
finite values of the Higgs bundle moduli.

In the following we would like to give a fairly general construction
of {\it algebraic} classes that satisfy $\alpha\cdot_{\bar C} \gamma
= 1$. We will apply it to $S={\bf P}^1\times {\bf P}^1$, but it
should be clear that with some simple substitutions it can also be
applied to the other cases. Thus we will finally establish some
examples of supersymmetric $SU(5)$ models with three generations and
$S$ contractible.

The strategy is as follows: we first take a curve $\alpha_0 \in
H^2(S,{\bf Z})$ such that $\alpha_0 \cdot_S \Sigma_\bt = 1$. Then we
will construct a curve $\alpha \in \bar X$ which does not intersect
$\Sigma_\bt$, and which covers $\alpha_0$ exactly once. Finally we
will require $\bar C$ to contain $\alpha$ by tuning the complex
structure moduli. The result is an algebraic class in $H^2(\bar C,
{\bf Z})$ with $\alpha\cdot_{\bar C} \gamma = 1$. We can then
construct an additional flux $\gamma_2$ as above by subtracting the
trace, and and define a spectral line bundle with
\be\label{newspectralL} c_1(L)\ =\ \half r + \half \gamma_u + n
\gamma_2 \ee
By adjusting $n$, we then get any number of generations we want.

In the case of interest, we have $S = {\bf P}^1 \times {\bf P}^1$.
We denote the coordinates on $S$ by $(z_1,z_2;w_1,w_2)$, and the two
rulings by $H_1$ and $H_2$, with intersection  numbers $H_1^2 =
H_2^2 = 0, H_1 \cdot H_2 =1$. As we deduced above, the matter curve
should be given by $\Sigma_\bt = [H_1 + H_2]$, and we need a class
with $\alpha_0 \cdot \Sigma_\bt = 1$. Thus a simple choice is to
pick $\alpha_0 = H_1$, though clearly there are additional options.
In equations it is given by (say) $w_1 = 0$.

Now we need to construct $\alpha$. As before we use coordinates
$(u_1,u_2)$ on the ${\bf P}^1$-fibers of $\bar X = {\bf P}(\cO
\oplus K_S)$. Then we define $\alpha$ by the following two
equations:
\be \alpha: \ w_1 = 0,  \qquad u_1 = P_2(z,w) u_2  \ee
where $P_2(z,w)$ is a section of $\cO(2,2)$. Note that $u_2 = 0
\Rightarrow u_1 \not = 0$, so $\alpha$ does not intersect the zero
section $u_2 = 0$. This is needed if we want $\bar C$ to contain
$\alpha$, because the intersection of $\bar C$ with $u_2=0$ is by
definition $\Sigma_\bt$, and we promised to construct a class which
does not intersect $\Sigma_\bt$. Also, $\alpha$ covers $\alpha_0$
precisely once.

Therefore we now need to show that we can tune the complex structure
moduli so that $\bar C$ contains $\alpha$. The equation of $\bar C$
is given by
\be\label{Cbareqn} b_0(z,w) u_2^5 + b_2(z,w) u_1^2 u_2^3 + \ldots +
b_5(z,w) u_1^5 = 0. \ee
We simply substitute the equations for $\alpha$ in order to get a
restriction on the coefficients of $\bar C$. Clearly we find that
\be w_1 \quad {\rm divides} \quad b_0(z,w) + b_2(z,w) P_2(z,w)^2 +
\ldots + b_5(z,w)P_2(z,w)^5 \ee
This can be satisfied by leaving $b_2, \ldots, b_5$ arbitrary, and
putting
\be\label{LNCbar} b_0(z,w) = -\left[b_2(z,w) P_2(z,w)^2 + \ldots +
b_5(z,w)P_2(z,w)^5\right]_{w_1 \to 0} + \cO(w_1) \ee
The only thing left to check is that $\bar C$ is generically smooth,
so that our calculations of the chiral spectrum apply. But this is
fairly obvious because generically the derivatives of equation
(\ref{Cbareqn}), even with (\ref{LNCbar}), give independent
equations.

Therefore we have constructed a $(1,1)$ class $\gamma_2 = 5\alpha -
p_C^*p_{C*}\alpha$ with the desired properties. Defining a spectral
line bundle as in (\ref{newspectralL}) with $n=8$, we find precisely
three chiral generations on $\Sigma_\bt$.

\newsubsection{`Flux' vacua in the heterotic string and $F$-theory}

As we already remarked, much of the structure of $F$-theory vacua is
identical with that of the heterotic string. BPS instantons effects,
branes and flux superpotentials, which are some of the main
ingredients inducing potentials for the moduli, can be related under
the duality. In particular, semi-realistic heterotic models appear
to have an enormous number of `flux' vacua as well. We put `flux' in
quotation marks here because after Fourier-Mukai transform, we get a
smooth non-abelian bundle on the Calabi-Yau three-fold without any
$U(1)$ fluxes. These extra `flux' vacua are obtained by using
spectral line bundles that are not inherited, and generically should
stabilize all vector bundle moduli. There is a landscape of such
vacua and a priori it is not clear why we should exclude these
possibilities. The method we used for constructing such more general
fluxes in section \ref{NoninheritedFluxes} can also be used in the
heterotic string.

Thus landscapes seem to be quite generic properties of
superpotentials in string theory. It would be interesting to study
these vacua microscopically. In the heterotic setting there are no
RR fields so one could try to use conventional CFT techniques.
Perhaps one may then find a reason to exclude most of them, although
it is currently not clear why that would be the case.

As in type II settings, this leads to philosophical problems: we
don't really understand moduli stabilization and the cosmological
constant problem very well, it is practically impossible to
enumerate all the vacua that seem to exist at the effective field
theory level, and one of the solutions that have been proposed to
solve the cosmological constant problem is NP hard
\cite{Denef:2006ad}. A possible way out was promoted in
\cite{Verlinde:2005jr}: if $M/M_{Pl}$ can be parametrically small,
where $M$ is some scale relevant for particle physics like the GUT
scale, then we can prevent the unknown physics responsible for
solving gravity-related problems from feeding back into physics at
the scale $M$. This may allow us to discuss phenomenology without
having to solve the cosmological constant problem and other problems
related to gravity. But combining this principle with GUTs leads us
to $F$-theory; this idea cannot be implemented in the heterotic
string.

\newsubsection{Conclusions}

We have clarified the rules for constructing local models in
$F$-theory. Such models can be defined by specifying suitable
spectral data (a type of $B$-brane) in an auxiliary Calabi-Yau
geometry. We classified the possible matter curves for local $SU(5)$
models. We have constructed the first truly local $SU(5)$ models
with three generations. It is still an open problem to construct
local $SU(5)$ models with {\it exactly} the MSSM spectrum, or some
acceptable extension. We explained how to connect $F$-theory models
to a IIb picture by taking an orientifold limit.

We also found that it seems to be impossible for a local $SU(5)$
model with completely unstabilized Higgs field moduli to have three
generations. From a physical perspective, this is good news since
there are many indications that we do not want a generic model, such
as problems associated to dimension four and five proton decay. Thus
requiring a three-generation model automatically stabilizes some of
the moduli. Requiring the precise MSSM spectrum will likely
stabilize even more moduli. On the other hand, this also makes the
problem of constructing realistic local models much more
challenging.

Along the way we have encountered a number of constraints that the
matter curves and fluxes on the matter curves must satisfy in a
consistent local model, from topological and integral (such as
anomaly cancellation (\ref{SU5tadpole}) and the fact that the flux
must lift to the integral class of a line bundle $L$ with
$c_1(p_{C*}L) = 0$), to analytic (such as the forced singularities
on $\Sigma_{\Lambda^2 E}$ and constraints on the moduli entering the
matter curves for non-inherited fluxes, so that $L$ is a holomorphic
line bundle). Probably we have not found them all, since some of
these constraints look very non-trivial from the point of view of
the brane carrying the unbroken gauge group. However we have shown
that there is an isomorphism relating a configuration $(E,\Phi)$ in
the $8d$ gauge theory to its spectral data $(C_E,L_E)$, and the
constraints on the spectral data are few and simple to understand.

In the next section we will make the first strides towards embedding
our local models in a global model.

 \newpage

\newsection{Compactification}
\seclabel{Compactification}

In order to get a finite four-dimensional Planck scale we should
embed our local models into a compact elliptically fibered CY
four-fold. In the philosophy of local model building, the goal of
this pursuit is not to find `the' UV completion of the local model.
Indeed as we reviewed earlier, it is not even clear that this is an
answerable question. Rather it is to ascertain that all the
ingredients used can in principle be combined in a UV complete
model, and there are no obvious constraints from UV completion that
would rule them out. In doing so there are many issues to be
addressed. Our aim here is rather modest; we would like to discuss a
simple set of compactifications which implement a few of the
requirements for viable local GUTs, and which make clear how such
constructions work in general. In particular we would like to
construct compactifications in which GUT breaking by fluxes can be
implemented, and in which dangerous proton decay channels can be
avoided.

Our discussion will have one important caveat. We will freely assume
that appropriate fluxes may be found which give precisely the
Standard Model spectrum on the matter curves we engineer. As we
emphasized in section 2, it has not yet been shown that this can
actually be done in a local model, let alone in a global model. The
point of our discussion is not to understand the fluxes, but rather
some of the constraints on the geometry of the four-fold arising
from phenomenological requirements.

\newsubsection{First example: cubic surface in ${\bf P}^3$.}

Let us discuss simple compactifications of local toy models with
$SU(5)$ GUT group. Our GUT brane should be wrapped on a Del Pezzo
surface $S_2 \subset B_3$, such that some homology classes in the
Del Pezzo become boundaries when embedded in $B_3$. For simplicity
we will take $B_3$ to be ${\bf P}^3$ in our first example, although
much of what we will say can clearly be adapted to more general Fano
three-folds. Then we can take $S_2$ to be a quadric surface
$Q_2(z_1,z_2,z_3,z_4)=0$ (i.e. ${\bf P}^1 \times {\bf P}^1$) or a
cubic surface $Q_3(z_1,z_2,z_3,z_4)=0$ (i.e. a Del Pezzo 6).  For
definiteness we take the cubic.

Recall again the Tate form of the Weierstrass equation
\be y^2 + a_1 xy + a_3 y = x^3 + a_2 x^2 + a_4 x + a_6 \ee
where the $a_i$ are sections of $K_{B_3}^{-i}$. In the case of $B_3
= {\bf P}^3$, the $a_i$ are polynomials of degree $4i$. As we
discussed in section, in order to get an $I_5$-locus along $Q_3 =
0$, as well as matter curves and Yukawa coupling localized along
certain prescribed submanifolds, we must impose certain restrictions
on the $a_i$ which can be read from table \ref{Tate}.

Now let us try to impose various constraints.
\begin{enumerate}
\item{\it $SU(5)$ gauge group on $Q_3=0$.} According to table \ref{Tate} this
implies the following leading form for the $a_i$:
\be a_1 = P_4, \qquad a_2 = s P_5, \quad a_3 = s^2 P_6, \qquad a_4 =
s^3 P_7, \qquad a_6 = s^5 P_9 \ee
where $s = Q_3$ and the $P_n$ are generic polynomials of degree $n$
on ${\bf P}^3$. They are identified with the sections $b_{9-n}$ that
appeared in the general discussion. In fact we are clearly allowed
to add further subleading terms, eg. $a_1 = P_4 + s T_1, \, a_2 = s
P_5 + s^2 T_2$, et cetera. Such additional subleading terms do not
affect behaviour near the $I_5$ locus, but they do provide
additional flexibility in building a global model. To keep things
simple, we will set them to zero. Then the discriminant is computed
to be
\be \Delta = s^5 P_4^4(-{P_9} {P_4}^2+{P_6} {P_7} {P_4}-{P_5}
   {P_6}^2) + \cO(s^6)\ee
which vanishes to 5th order along $Q_3=0$, as required.

\item{\it Matter curves.} The discriminant vanishes to higher order along $\Sigma_\bt = \{
Q_3 = P_4 = 0 \}$ and $\Sigma_\bfv = \{Q_3 = R_{17}=0\}$, where
\be R_{17} \equiv{P_9} {P_4}^2-{P_6} {P_7} {P_4}+{P_5}{P_6}^2. \ee
Note that
\be \Lambda^2T_{Q_3} = \cO(1)|_{Q_3}, \qquad N_{Q_3} = \cO(3)|_{Q_3}
\ee
and hence the cohomology classes dual to $\Sigma_\bt$ and
$\Sigma_\bfv$ on $Q_3$ are given by $c_1 - t$ and $8 c_1 - 3t$. Of
course this all fits in the general discussion in sections
\ref{LocalfromGlobal} and \ref{Tadpoles}.

\item{\it Yukawa couplings and dimension four proton decay.}
The up-type Yukawa couplings are localized at $\{Q_3=P_4 = P_5 =
0\}$ and the down type Yukawas are localized at $\{Q_3 = P_4 = P_6 =
0\}$. Methods for suppressing proton decay in $F$-theory were
discussed in \cite{Tatar:2006dc,Donagi:2008kj,Beasley:2008kw}. Here
we will see how they can be implemented in global models.

In order to prevent dimension four proton decay, we want to make
sure that
\be \begin{array}{l} \bt_m \cdot \bfb_m \cdot \bfb_m \\[2mm] \bt_m \cdot
\bfb_h \cdot \bfb_h
\end{array}
\Rightarrow {\rm absent},
\qquad \bt_m \cdot \bfb_m \cdot \bfb_{h_d} \Rightarrow {\rm present}
\ee
This can be done by splitting $\Sigma_\bfv$ into two pieces, one
supporting the matter fields and another supporting the Higgses.
This means we have to tune the $P_n$ so that the polynomial $R$
factorizes modulo $Q_3$. In terms of ideals, we require a
decomposition
\be \vev{Q_3,R} = I_{\bfb_m} \cap I_h \ee
To secure the absence of $R$-parity violating down-type Yukawa
couplings we must make sure that whenever we have an intersection of
$\Sigma_{\bt_m}$ with $\Sigma_{\bfb_m}$, there is also a branch of
$\Sigma_{\bfb_h}$ intersecting at that point. Since $R=0$ has a
double point at such intersections, we can also say that whenever
$\Sigma_{\bt_m}$ intersects with $\Sigma_{\bfb_m}$, then
$\Sigma_{\bfb_m}$ is not allowed to have a double point (the second
order vanishing of $R$ instead being due to a branch of
$\Sigma_{\bfb_h}$ coming in and intersecting there). Similarly in
order to avoid the couplings $\bt_m \cdot \bfb_h \cdot \bfb_h$, we
want to avoid double points on $\Sigma_{\bfb_{h_d}}$ which also meet
$\Sigma_{\bt_m}$.

We don't know the general solution to this algebraic problem. But to
see that it can be achieved, we will exhibit one simple solution
that exists for more general $SU(5)$ models as well. We take
\be
\begin{array}{rclrcl}
  P_6 &=& H^d_6 \ \ {\rm mod\ } Q_3 \qquad &P_9 &=& H^u_1 T_2 H^d_6\ {\rm
mod\ } Q_3 \\
  P_7 &=& H^u_1 T_6 \ {\rm mod\ } Q_3 \qquad & P_5 &=&
H^u_1 T_4 \ \ \ {\rm mod\ } Q_3
\end{array}
\ee
for some $T_i$ and $H_i$ of the appropriate degree, but otherwise
arbitrary. Then we take $\Sigma_{\bfv_h} = \{Q_3 = H^u_1 H^d_6 =
0\}$, i.e. the Higgs curve is actually reducible, with only up type
Yukawa couplings on $H^u_1 = 0$ (since $H^u_1 = 0$ implies $P_5 =
0$) and only down type Yukawa couplings on $H^d_6=0$. When we
discuss dimension five proton decay we will see why that is a good
thing to have. Now we can factorise $\Sigma_\bfv$ as
\be R_{17} = (H^u_1 H^d_6)  \cdot  M_{10} \ {\rm mod\ } Q_3,\qquad
M_{10} = T_2 P_4^2 - T_6 P_4 + T_4 P_6 \ee
and $M_{10}$ has no double points at $Q_3 = P_4 = P_6 = 0$. Moreover
the up and down-type Yukawa's are still present. For instance the
up-type Yukawa's come from $Q_3 = P_4 = H^u_1 = 0$, which consists
of $3 \cdot 4 \cdot 1 = 12$ points.

There are additional cubic couplings of the form
\be\label{55bcouplings} \bfb_{m} \cdot \bfv_{h_u} \cdot {\bf 1},
\qquad \bfb_{h_d} \cdot \bfv_{h_u} \cdot {\bf 1} \ee
The singlets correspond to Higgs field moduli (which are complex
structure moduli of the Calabi-Yau four-fold). At least three of
them should give rise to right-handed neutrinos, with the first
coupling in (\ref{55bcouplings}) corresponding to the usual Yukawa
couplings for neutrinos. The number of moduli appearing in such
couplings is the difference between the number of moduli describing
$\Sigma_{\bfb_m}$ and $\Sigma_{\bfv_{h_u}}$ separately or as a
single smooth curve, which yields 65 singlets in our example. The
problem of getting Majorana masses of the right order of magnitude
is a problem of moduli stabilization for the Higgs fields. The
couplings on the right give rise to the minimal extension of the
MSSM with a dynamical $\mu$-parameter. There are additional
constraints from dimension five proton decay however, as we discuss
next.

\item{\it Dimension five proton decay.}
We further want to eliminate dimension five proton decay. This
proceeds through mediation of massive KK triplets $T_u,T_d$
propagating on curves supporting a hypermultiplet in the $\bfv$. The
possible channels are given by
\be\label{dim5decay}
\begin{array}{ccccccc}
 Q\, Q                    & \stackrel{\lambda_u}{\longleftrightarrow}& T_u          &
 \stackrel{m^{ab}}{ \longleftrightarrow}& T_d         & \stackrel{\lambda_d}{\longleftrightarrow}& Q\, L \eol [2mm]
\Sigma_{\bt_m} \times \Sigma_{\bt_m} &  &  \Sigma_\bfv^a & &
\Sigma_\bfv^b & & \Sigma_{\bt_m} \times \Sigma_{\bfb_m}
\end{array}
\ee
In order to prevent such processes, we have to shut off at least one
of the interactions in this chain. If  $H_u$ and $H_d$ propagate on
the same matter curve, and if we assume the existence of classical
up-type and down-type Yukawa couplings for the Standard Model, then
such decays are unavoidable. Since we want to keep the classical
Yukawa couplings, we require the existence of a decomposition
\be I_h = I_u \cap I_d \ee
%
%
so that we can shut off the coupling $m^{ab}$. If we allow
$\Sigma_u$ and $\Sigma_d$ to intersect, then there could either be a
branch of $\Sigma_{\bt_m}$ also intersecting there;
or it can correspond to a $\bfv_u \cdot \bfb_d \cdot {\bf 1}$
coupling. As long as the VEV of the singlet vanishes we do not have
the troublesome mass terms linking triplets localized on $\Sigma_u$
and $\Sigma_d$. In either case the existence of a classical
$\mu$-term is excluded. Our example corresponds to the latter case:
there are $3\cdot 1 \cdot 6=18$ intersection points on $H^u_1 =
H^d_6 = 0$ corresponding to the couplings $\bfv_u \cdot \bfb_d \cdot
{\bf 1}$.

There are several possible alternate channels for dimension five
proton decay (\ref{dim5decay}). The most dangerous are cases where
$\Sigma^a_\bfv = \Sigma^b_\bfv$, because then mass terms $m^{ab}$
between $T_u$ and $T_d$ cannot be avoided. The case $\Sigma^a_\bfv =
\Sigma^b_\bfv = \Sigma_{\bfb_m}$ is harmless by the solution to
dimension four proton decay, which shuts off the interactions
$\lambda_d$. The case $\Sigma^a_\bfv = \Sigma^b_\bfv =
\Sigma_{\bfb_{h_d}}$ requires shutting off the interactions
$\lambda_u$. The curve $\Sigma_{\bt_m}$ is positive in our example
and therefore certainly intersects $\Sigma_{\bfb_{h_d}}$. However in
our solution to the dimension four problem, by design any such
intersection has $P_4 = P_6 = 0$ and therefore corresponds to a
$\lambda_d$ coupling, not a $\lambda_u$ coupling, so this channel is
not available. Finally there is the case $\Sigma^a_\bfv =
\Sigma^b_\bfv = \Sigma_{\bfv_{h_u}}$, which requires shutting off
the interaction $\lambda_d$. In our example, by design any
intersection point between $\Sigma_{\bfv_{h_u}}$ and $\Sigma_\bt$
yields an up-type Yukawa, so the potentially troublesome
interactions are again absent.

The remaining possible channels have $\Sigma_\bfv^a \not =
\Sigma_\bfv^b$. Assuming both the $\lambda_u$ and $\lambda_d$
couplings are present (which they need not necessarily be), the
problem is to shut off the interactions $m^{ab}$. This depends on
the existence of intersections of $\Sigma_\bfv^a$ and
$\Sigma_\bfv^b$ which give rise to a coupling $\bfv \cdot \bfb \cdot
{\bf 1}$. If such intersections are present, $m^{ab}$ is
proportional to the VEV of the singlet, which is a complex structure
modulus. As long as the dynamics of moduli stabilization is such
that the VEV of this field remains zero, there will be no proton
decay through this channel.

Consider for instance $\Sigma_\bfv^a =\Sigma_{\bfb_m}$ and
$\Sigma^b_\bfv =\Sigma_d$ (the other cases being similar). The
curves $\Sigma_{\bfb_m}$ and $\Sigma_d$ can intersect in two ways.
Either there is also a branch of $\Sigma_\bt$ intersecting there,
which corresponds to the down type Yukawa's that we want to have; or
it corresponds to a $\bfv \cdot \bfb \cdot {\bf 1}$ coupling.  In
our example with $I_{h_d} = \vev{Q_3,H^d_6}$, intersection points
where $H^d_6 = 0$ and $M_{10} = 0$ have either $P_4 = 0$ or $T_2 P_4
- T_6 = 0$. In the former case it meets with $\Sigma_\bt$,  and
there are $3 \cdot 6 \cdot 4=72$ such intersection points; in the
latter case it corresponds to the coupling to singlets whose VEV
must remain zero, and this accounts for $3 \cdot 6 \cdot 6 = 108$
intersection points.

\end{enumerate}

Hence we see in this simple example that there is enough room in
complex structure moduli space to implement our geometric
requirements for absence of dimension four and five proton decay. In
fact although we did not write the most general solution above, it
seems a solution along these lines is required. In order to
eliminate the double points from $\Sigma_{\bfb_m}$ and
$\Sigma_{\bfb_{h_d}}$, we should factor out $P_6$ from $R$, and in
order to avoid dimension five proton decay, we should make sure that
$\Sigma_{\bfb_{h_d}}$ is contained in $P_6 = 0$ and
$\Sigma_{\bfv_{h_u}}$ is contained in $P_5 = 0$.

The solution we provided though required the VEVs of certain complex
structure moduli to remain vanishing. This is a requirement we must
impose on the moduli stabilization mechanism, which we have not
considered here, and on the face of it does not seem particularly
natural (although it is technically natural). One might speculate
there are extra supersymmetric fluxes available for these values of
the moduli which we should turn on in order to recover the precise
Standard Model spectrum. That would be a nice way to really explain
moduli stabilization and lack of proton decay in our models, but it
seems currently unclear why that should be the case. An alternative
approach would be to ensure that the potentially troublesome
intersection points are all absent, which seems much harder to
arrange, or to implement the approach of \cite{Tatar:2006dc}, which
requires decomposing the $SU(5)$ Casimirs into those of a smaller
holonomy group and then making a small deformation, so that one has
additional $U(1)$'s available.

\newsubsection{Second example: a contractible ${\bf P}^1 \times {\bf P}^1$.}

In our previous example the del Pezzo was not contractible in $B_3$.
The main purpose of this subsection is to give a simple example of a
del Pezzo $S$ which has two-cycles not inherited from $B_3$
(necessary for allowing GUT breaking fluxes), and which is also
contractible in $B_3$. This is an explicit realization of case
(\ref{goodmodel}) discussed in section \ref{NoGo}.

The example is as follows. We will take $B_3$ to be the blow-up of
${\bf P}^3$ along a curve $C$ defined by
\be\label{Cdefn} C = \{ Q_2=0\} \cap \{Q_3 = 0\} \ee
The corresponding
ideal is denoted as $I_C$ and the blow-up along this ideal as $B_3 =
\tilde{\bf P}^3$. We have
\be {K}_{\tilde{\bf P}^3} = i^*{K}_{{\bf P}^3} + \tilde{C} \ee
where $\tilde{C}$ is the exceptional divisor (a ${\bf
P}^1$-fibration over $C$, given by projectivising the normal
bundle). Sections of the anti-canonical bundle ${K}_{\tilde{\bf
P}^3}^{-1}$ are sections of ${K}_{{\bf P}^3}^{-1}$ which are also in
the ideal $I_C$. In particular there are non-trivial sections in
${K}_{\tilde{\bf P}^3}^{-4}$ and ${K}_{\tilde{\bf P}^3}^{-6}$, and
so we can write a Weierstrass equation and construct elliptic
fibrations over $\tilde{\bf P}^3$ which are Calabi-Yau.

In this example, the del Pezzo on which the gauge branes are wrapped
will be the Hirzebruch surface ${\bf F}_0={\bf P}^1 \times {\bf
P}^1$, here defined by $Q_2 = 0$. As mentioned, the reason for
picking this model as our next example is that the proper transform
of $S = \{Q_2 = 0\}$ is contractible. To see this, let us first
check that the normal bundle is indeed negative. The normal bundle
of $S$ in ${\bf P}^3$ is $\cO(2)|_S$. After blowing up $C$, the new
normal bundle is $\cO(2)|_S \otimes \cO(-\tilde{C})|_S$, where
$\tilde C$ is the exceptional divisor. But the intersection of
$\tilde C$ with $S$ is in $\cO(3)|_S$, so
\be\label{newnormal} \cO(2)\otimes \cO(-\tilde{C})|_S \sim
\cO(-1)|_S. \ee
Hence the normal bundle is negative, a necessary condition for being
contractible.

By looking at this example slightly differently, one may establish
that $S$ is indeed contractible. Consider a cubic hypersurface $Q$
in ${\bf P}^4$ vanishing to second order at a point $p$. Let $T$ be
the tangent space to ${\bf P}^4$ at $p$. We identify $T$ with an
open subset of ${\bf P}^4$, and write the Taylor expansion of $Q$ at
$p$ as:
\be Q = Q_2 +Q_3, \ee
with $Q_2$, $Q_3$ as in (\ref{Cdefn}). If we also identify the ${\bf
P}^3$ of (\ref{Cdefn}) with the projectivization of this $T$, we see
that the set of lines in $Q$ through its singular point $p$ can be
identified with the curve $C$.

Consider the projection $\widetilde{{\bf P}^4} \to {\bf P}^3$ with
center $p$, where $\widetilde{{\bf P}^4}$ is the blowup of ${\bf
P}^4$ at $p$. It restricts to a surjective morphism $\pi:
\widetilde{Q} \to {\bf P}^3$, where $\widetilde{Q}$ is the blowup of
$Q$ at $p$. The exceptional divisor $S_0$ in $\widetilde{Q}$ is
mapped by $\pi$ isomorphically to a quadric surface in ${\bf P}^3$
that can be identified with our surface $S$. On the other hand, the
inverse image of each point of $C \subset {\bf P}^3$ is the
corresponding line in $Q$. We thus have an identification of
$\widetilde{Q}$ with $B_3 = \widetilde{{\bf P}^3}$, showing that $S$
can indeed be blown down to the singular point $p$ of the cubic
threefold $Q$.

In order to break the GUT group without generating a mass for
hypercharge, we need a class in $S$ which is topologically trivial
in $B_3$. For $S = {\bf P}^1 \times {\bf P}^1$ there is a unique
candidate, the difference between the two rulings. It's not hard to
see that the two ${\bf P}^1$'s yield equivalent classes in $B_3$:
they have the same intersection number with the transform of the
hyperplane class in ${\bf P}^3$, as well as with the exceptional
divisor.

Now let us write explicitly the elliptic fibration. Once again we
recall the Tate form of the Weierstrass equation
\be y^2 + a_1 xy + a_3 y = x^3 + a_2 x^2 + a_4 x + a_6 \ee
and repeat the exercise of the previous subsection. The $a_i$ are
sections of ${K}_{\tilde{\bf P}^3}^{-i}$, which means they are
polynomials of degree $4i$ on ${\bf P}^3$ which vanish to $i$th
order along $C$. Let us define $s = Q_2$ and $u = Q_3$. A polynomial
which vanishes to $i$th order along $C$ is a sum of terms each of
which is of degree at least $i$ in $s$ and $u$.

\begin{enumerate}
\item{\it $SU(5)$ gauge group on $Q_2=0$.} According to table \ref{Tate} this
implies the following form for the $a_i$:
\be
\begin{array}{rclrcl}
  a_1 & = & P_1 u + P_2 s \qquad  & a_4 & = & s^3 (P_7u +P_8 s) \\
  a_2 & = & s(P_3 u + P_4 s) & a_6 & = & s^5
(P_{11} u + P_{12} s)  \\
  a_3 & = & s^2 (P_5 u + P_6 s) \qquad &  & &
\end{array}
\ee
Here we included various subleading terms ($P_2, P_4, P_6,P_8,
P_{12}$) which are not directly needed, and which will not show up
in the analysis of the matter curves and Yukawa couplings.

\item{\it Matter curves.}
The discriminant is given by
\be \Delta = s^5 u^7 P_1^4 R_{13} + \cO(s^6) \ee
where
\be R_{13} = -P_1^2 P_{11} + P_1 P_5 P_7 - P_3 P_5^2 \ee
If we recall that $\tilde C$ is a ${\bf P}^1$-fibration over $C$,
then the intersection of $\tilde C$ with $s=0$ gives the section at
`zero' and with $u=0$ gives the section at `infinity.' Hence after
blowing up along $C$, the surfaces $s=0$ and $u=0$ no longer
intersect; instead they intersect $\tilde C$ along two disjoint
curves. Thus from the discriminant we read off that the matter
curves are given by
\be \Sigma_\bt = \{Q_2 = P_1 = 0 \} \ee
which is generically a rational curve, and
\be \Sigma_\bfv = \{Q_2 = R_{13} = 0\}. \ee
Recall we showed above that $N_S = \cO(-1)|_S$, and it is not hard
to see that $c_1 \sim \cO(2)|_S$. Therefore the homology classes of
the matter curves are given by $ \cO(1)|_S\sim c_1 - t $ and $
\cO(13)|_S\sim 8c_1 - 3t$, in full agreement with the general
discussion.

\item{\it Yukawa couplings and proton decay.} As is familiar by now,
the Yukawa couplings are localized at $\lambda_{up} \sim \{ Q_2 =
P_1 = P_3 = 0\}$ and $\lambda_{down}\sim \{Q_2 = P_1 = P_5 = 0\}$.
The discussion of the first example goes through if we choose the
analogous factorization:
\be
 P_{11} = P_5 T_5 H^u_1\quad {\rm mod}\ Q_2, \qquad
 P_7 = T_6 H^u_1 \quad {\rm mod}\ Q_2, \qquad
 P_3 = T_2 H^u_1 \quad {\rm mod}\ Q_2.
 \ee
With this factorization we have
\be R_{13} = (P_5 H^u_1) \, M_{7} , \qquad M_7 = -P_1^2 T_5 + P_1
T_6 - T_2 P_5 \ee
and we identify $\Sigma_{\bfb_{h_d}}= \{Q_2 = P_5 = 0\}$,
$\Sigma_{\bfv_{h_u}}= \{Q_2 = H^u_1 = 0\}$, and
$\Sigma_{\bfb_m}=\{Q_2 = M_7 = 0\}$. We refer to the discussion in
the first example for why this eliminates the classical dimension
four and five proton decay.

\end{enumerate}

As we saw, in local models of this type it is possible to engineer
three generations by putting a mild restriction on $P_{11}$. Whether
there exist fluxes which yield the Standard Model spectrum, and
whether these fluxes can be extended globally is an open question.

This example is really a special case of a more general
construction. Consider blowing up a Fano three-fold along a curve
$C$, and assume that the blow-up still admits a CY $T^2$-fibration.
The proper transform of the surface $S$ of minimal degree containing
$C$ is usually contractible, by the reasoning around equation
(\ref{newnormal}). Moreover such a surface will typically have
homology classes which are not inherited from the ambient space, as
in the example above, if the surface had such classes before blowing
up. If the degree of the surface is not too large, we can prescribe
an $I_5$ fibration along it.

\bigskip

\noindent {\it Acknowledgements}:

We would like to thank V.~Alexeev, F.~Denef, M.~Gross and T.~Pantev
for helpful discussions, and J.~Heckman for patiently explaining
many details of his works
\cite{Beasley:2008dc,Beasley:2008kw,Heckman:2008qa}. The research of
R.D. is supported by NSF grant DMS 0612992 and Research and Training
Grant DMS 0636606.

\newpage

\appendix

\renewcommand{\newsection}[1]{
\addtocounter{section}{1} \setcounter{equation}{0}
\setcounter{subsection}{0} \addcontentsline{toc}{section}{\protect
\numberline{\Alph{section}}{{\rm #1}}} \vglue .6cm \pagebreak[3]
\noindent{\bf Appendix {\Alph{section}}:
#1}\nopagebreak[4]\par\vskip .3cm}

\end{document}